\documentclass[aps, prb, twocolumn, superscriptaddress, english]{revtex4-2}

\usepackage{graphicx}
\usepackage{cancel}
\usepackage{enumerate}
\usepackage{hyperref}
\hypersetup{
colorlinks=true,
citecolor=blue,
linkcolor=blue,
urlcolor=blue,
}
\usepackage{textcomp}
\usepackage{amsmath}
\usepackage{amssymb}
\usepackage{pgf}
\usepackage{soul}
\usepackage{subfigure}
\usepackage[english]{babel}
\usepackage[normalem]{ulem}

\def\ee{\mathrm{e}}

\newcommand{\f}{\frac}
\newcommand{\rv}{\mathbf{r}}
\newcommand{\rvp}{\mathbf{r}^{\prime}}

\graphicspath{{./}{./Figure/}{./figures}}

\begin{document}
\title{Cooper quartets and fractional vortices in frustrated Josephson junction dice arrays}

\author{Erik Lennart Weerda}
\affiliation{Institute for Theoretical Physics, University of Cologne, D-50937 K\"{o}ln, Germany}
\affiliation{Forschungszentrum J\"{u}lich GmbH, Institute of Quantum Control,
Peter Gr\"{u}nberg Institut (PGI-8), 52425 J\"{u}lich, Germany}

\author{Olav F. Syljuåsen}
\affiliation{Department of Physics, University of Oslo, P. O. Box 1048 Blindern, N-0316 Oslo, Norway}

\author{Matteo Rizzi}
\affiliation{Institute for Theoretical Physics, University of Cologne, D-50937 K\"{o}ln, Germany}
\affiliation{Forschungszentrum J\"{u}lich GmbH, Institute of Quantum Control,
Peter Gr\"{u}nberg Institut (PGI-8), 52425 J\"{u}lich, Germany}

\author{Michele Burrello}
\affiliation{Dipartimento di Fisica, Università di Pisa, and INFN, Sezione di Pisa, Largo Pontecorvo 3, I-56127 Pisa, Italy}

\date{\today}

\begin{abstract}

Superconductivity mediated by Cooper quartets of charge $4e$ is a phenomenon of key importance for both the understanding of certain exotic superconductors and the engineering of quantum memories protected at the hardware level by topological order. Josephson junction arrays in the shape of the dice lattice constitute one of the main candidates for the realization of this phase of matter.

Here, we analyze numerical signatures of the emergence of this exotic phase when superconducting dice arrays are frustrated by inserting one third of a flux quantum per rhombic plaquette.

We adopt simulations of relaxation dynamics to study the critical current of such devices and analyse the Fourier decomposition of their bulk supercurrents. A further characterization of these systems at finite temperature is obtained through Monte Carlo techniques and two-dimensional infinite tensor networks. Our results indicate that the observed peaks of the critical current and temperature at frustration $1/3$ correspond to a superconductor-insulator phase transition compatible with the deconfinement of half-vortices, while the low-temperature correlation functions of the model confirm the onset of a $4e$ superconducting phase mediated by Cooper quartets.

We finally address the effects of Josephson energy and flux disorder typical of experimental arrays, and comment on the role of charging energies in the corresponding two-dimensional quantum model.

\end{abstract}

\maketitle

\section{Introduction}

In the study of exotic superconducting phases of matter, the investigation of models where superconductivity is not mediated by Cooper pairs is a recurring theme. In particular, superconductivity obtained via the condensation of Cooper multiplets, for instance pairs of Cooper pairs with charge $4e$, has been at the focus of a long-standing research endeavor stemming from the analysis of models which combine superfluidity and magnetic effects \cite{korshunov1985,kivelson1990}.
The condensation of $4e$ carriers has been considered in several contexts: On one side, it was proposed as a possible mechanism to describe exotic superfluid states arising in cuprates, as a result of the thermal melting of pair density wave states \cite{Berg2009}, and it plays a role in the modelling of nematic superconductors \cite{Jian2021, Fernandes2021}; on the other, it is conjectured to appear on a mesoscopic level through the engineering of suitable frustrated superconducting arrays \cite{Abilio1999} and it is considered a preliminary step for the realization of states with topological order \cite{Ioffe2002,Doucot2003}.

These artificial realizations of the condensation of $4e$ Cooper quartets have faced a resurgence of interest in the very last years, thanks to the development of scalable and electrically tunable Josephson junction arrays based on hybrid superconductor-semiconductor systems \cite{bottcher2018,bottcher2022,Bottcher2022b,Sasmal2025,Bondar2025,wang2026}. Hybrid Josephson junctions offer indeed the possibility of controlling the amplitude of Cooper pair tunneling between neighboring superconducting islands by varying the density of charge carriers in the semiconductor environment \cite{Shabani2016,Kjaergaard2017,Casparis_NatNanoTech2018,Ciaccia2023,Banszerus2024}. This allows, in turn, for controlling the average Josephson energy of the arrays through the use of global electrostatic gates \cite{bottcher2018,bottcher2022,Bottcher2022b,Sasmal2025,Bondar2025}.

The transport properties of these superconducting arrays provide fundamental information to characterize their states as a function of various parameters such as temperature, applied current bias, or external magnetic field. Their resistivity, for instance, allows for the observation of the Berezinskii-Kosterlitz-Thouless (BKT) transitions that typically separate the superconducting and normal states of these systems \cite{Fazio2001,Newrock2000}.

\begin{figure}
\includegraphics[width=\linewidth]{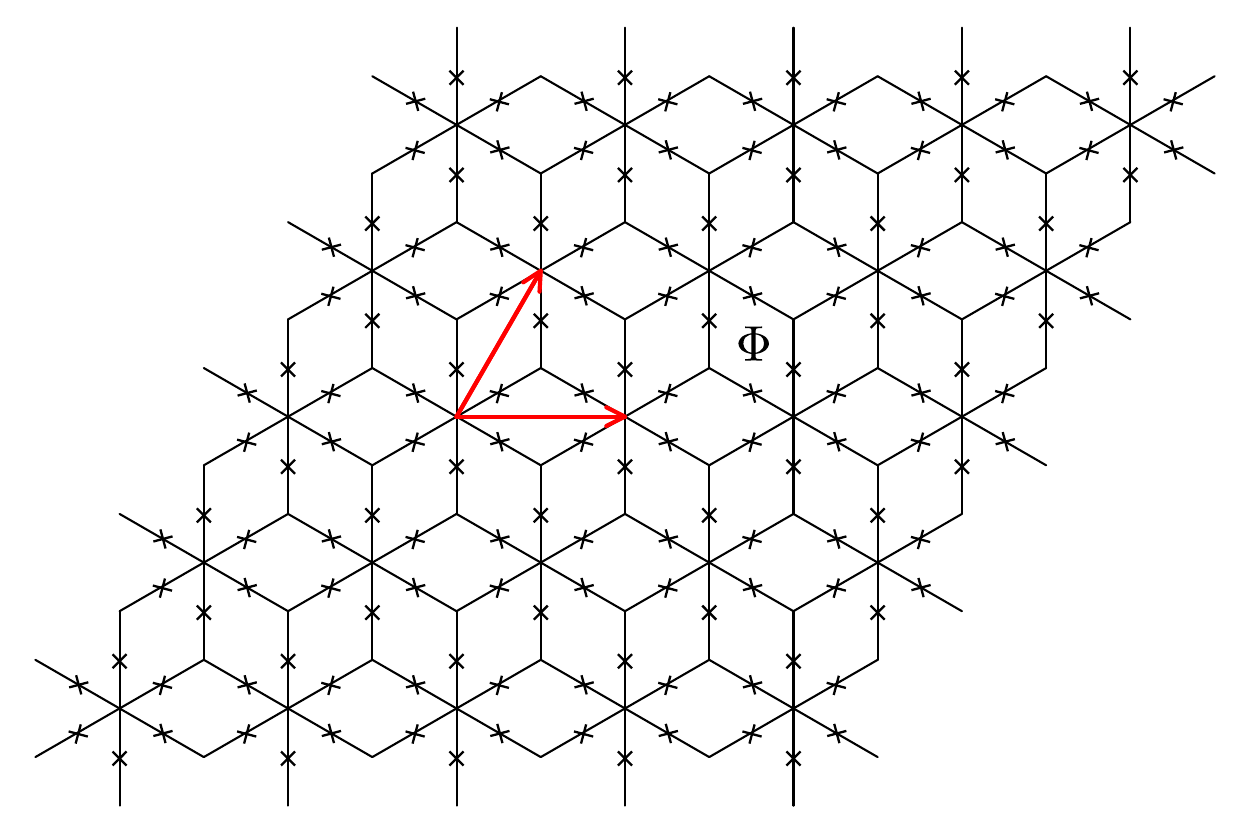}
\caption{
Illustration of a JJA in dice lattice geometry. The red arrows show the unit-cell vectors. By applying a perpendicular magnetic field to the array, a magnetic flux $\Phi$ can be threaded through every plaquette.}
\label{Fig:dice}
\end{figure}

In this work, we theoretically address magnetically frustrated Josephson junction arrays (JJAs) composing a dice lattice geometry (Fig.~\ref{Fig:dice}). This lattice has been broadly studied due to the onset of flat bands in the related particle dispersion \cite{Sutherland1986,Vidal1998,Rizzi2006}, and it is known for the extensive degeneracy of its low-energy vortex patterns at specific values of the magnetic flux per plaquette \cite{Korshunov2004,Korshunov2005,Korshunov2005b}.
We analyze the low-energy physics of the dice lattice as a function of its frustration and, based on several numerical methods, we show the onset of a condensed \textit{$4e$ phase} of Cooper quartets when the magnetic field corresponds to a flux $\Phi_0/3$ per rhombic plaquette.

Achieving a $4e$ superconducting phase acquires a special importance as this phase can be considered a classical precursor of states with topological order \cite{Ioffe2002,Doucot2003}, since half-vortices are topological excitations in the same topological class of the magnetic excitations in Kitaev's surface codes \cite{Kitaev2003} and $\mathbb{Z}_2$ gauge theories \cite{Senthil2000}. Indeed, these quantum models can be interpreted also as $4e$ superfluids whose topological excitations are the charge $2e$ Cooper pairs and the magnetic half-vortices. Their topological nature is embodied by the Aharonov-Bohm phase $\ee^{i\pi}$ acquired by a Cooper pair encircling a half-vortex.

The paradigmatic example for the $4e$ phase is represented by the Lee-Grinstein model \cite{Lee1985}, which extends the conventional XY-model by introducing an additional $\pi\text{-periodic}$ interaction between neighboring sites. This model displays indeed two different superconducting phases corresponding to a standard Cooper pair superfluid and a $4e$ superfluid. Its intricate phase structure is related to two key elements: (I) the presence of \textit{half-vortex excitations}, characterized by a fractional vorticity which, translated into the physical superconducting arrays, corresponds to a magnetic flux $\Phi_0/2 = h/4e$; (II) the weak string tension of the domain walls which connect pairs of half-vortices. Indeed, if the string tension of these domain walls vanishes at a temperature $T^*$ lower than the BKT temperature $T_{\rm BKT}$ associated with the half-vortices, then the $4e$ phase emerges at intermediate temperatures $T^* < T < T_{\rm BKT}$, before the eventual deconfinement of the half-vortices suppresses the superconducting stiffness and the critical current.
While the Lee-Grinstein model gives an intuitive picture for the emergence of $4e$ superfluids, the considered $\pi$-periodic interactions are not straightforwardly realizable in large JJAs. 

Korshunov, however, conjectured that very similar physics characterize the pure XY-model describing conventional JJAs on the dice lattice both at full frustration, where half a quantum of flux is threaded through every plaquette, $f = \Phi/\Phi_0 =1/2$ \cite{Korshunov2005}, and at $f=1/3$ \cite{Korshunov2005b}. 
For both values of the magnetic frustration in the XY dice model, fractional vortices are associated with the intersection of zero-energy domain walls, which display a vanishing string tension at $T^*=0$. Nevertheless, there are differences between these two cases. For $f=1/2$, the fractional vortices emerging in the system are conjectured to be characterized by fluxes smaller than $\Phi_0/2$ ($\Phi_0/8$ and multiples) \cite{Korshunov2005}. This implies a lower  critical temperature $T_{\rm BKT}$ and critical current, as their value is suppressed by the reduced vorticity based on renormalization group arguments \cite{Korshunov2005}. 
At $f=1/3$, instead, the fractional vortices are half-vortices associated with the merging of three domain walls in a single hexagon of the lattice. Given their flux $\Phi_0/2$, the critical temperature $T_{\rm BKT}$ is expected to be higher than the $f=1/2$ case and approximately given by one quarter of the critical temperature of the BKT transition in the absence of external magnetic field.

Motivated by recent experiments in JJAs with a dice lattice geometry \cite{Bondar2025}, we present in the following a theoretical study of several properties of the related XY model as a function of magnetic frustration. In Sec.~\ref{sec:general} we analyze the expected landscape of critical currents as a function of the magnetic frustration $f$, which displays characteristic features at $f=1/6, 1/3$ and $1/2$. We consider additional transport features to evaluate the contribution of Cooper quartets to the supercurrents in the system and we observe that their role is predominant for $f=1/3$.

In Sec. \ref{sec:13} we focus on the case $f=1/3$ and discuss the emergence of a $4e$ phase at low temperatures. 
Sec. \ref{sec:charging} analyzes the role of charging energy and quantum fluctuations in these superconducting arrays and discusses the possibility of order-by-disorder phenomena that may cause a transition from the $4e$ phase to ordered vortex-lattice states at extremely low temperatures.

\section{Features of the supercurrents in the frustrated dice lattice} \label{sec:general}

The recent developments of lithographic techniques for the fabrication of hybrid superconductor-semiconductor devices enabled the realization of large JJAs (systems with more than $10^4$ superconducting islands) characterized by good control of the average Josephson energy of their constituting junctions \cite{Bondar2025,bottcher2018,bottcher2022,Bottcher2022b,Sasmal2025,Reinhardt2025}. 

The typical lattice size for these arrays is of order $0.1 - 1$\textmu m, and the precision of their fabrication is such that the expected average disorder in the areas of the plaquettes and thus in the magnetic flux per plaquette, is below $3\%$. The disorder in the distribution of the Josephson energies, instead, is estimated around $10 \%$ (see, for instance, Ref. \cite{manucharyan2019}).
Furthermore, for superconducting island sizes of order $1$\textmu m$^2$ the effects of electrostatic interactions, thus quantum fluctuations, appear negligible \cite{bottcher2018,bottcher2022,Bottcher2022b}, such that a classical description of the superconducting arrays captures their main many-body features.

In these hybrid superconducting arrays, the expected phenomenology of the XY models is recovered for setups in which the inner part of the semiconductor plaquettes is depleted through the use of suitable global electrostatic gates \cite{Bondar2025}. In such a situation, the physics of the hybrid Josephson junction arrays approaches the behavior of thin superconductor networks \cite{Abilio1999,Serret2002,Tesei2006}. The intricate pattern of superconducting switching currents, in particular, becomes periodic with the frustration parameter $f$ \cite{Bondar2025}. This indicates that transport occurs through the inter-island junctions only, whereas the contribution of the semiconductor plaquettes vanishes as suggested by the suppression of the related interferometric effects \cite{Bondar2025}.

Therefore, the low-energy behavior of the dice JJAs with depleted semiconductor plaquettes can be efficiently captured by a theoretical description based on the XY model and it displays most of the common features expected from standard 2D superconducting arrays \cite{Bondar2025,Baek2008}: They are characterized by a low-temperature superconducting phase and a high-temperature normal phase with non-vanishing resistance, such that transport measurements through these systems yield direct observations of their critical temperatures as a function of the magnetic frustration $f$. The superconducting phase can also be broken by biasing the full superconducting array with a strong current flowing through the whole lattice. In this case, the system remains superconducting up to a critical switching current bias $I_c$.

\subsection{The critical current in the frustrated dice lattice} \label{sec:criticalcurrent}

The superconducting arrays we consider are characterized by islands with negligible charging energies, such that their dynamics can be modelled through a classical 2D XY model \cite{Newrock2000}
\begin{equation}
H= -E_J \sum_{\left\langle{\bf r},{\bf r'} \right\rangle} \cos \left[\varphi_{\bf r} - \varphi_{\bf r'} -A_{\bf r r'} \right]\,,
\label{eq:2dXYmodel}
\end{equation}
where $E_J$ is the Josephson energy of each junction in the lattice and the sum is taken over nearest neighbor islands on the dice lattice. We neglect the dependence of $E_J$ on temperature and magnetic field.
The Peierls phases 
$A_{\bf r r'}$
are such that their oriented sum along any rhombic plaquette $p$ of the lattice returns
\begin{equation}
\sum_{\left\langle{\bf r},{\bf r'} \right\rangle \in p} A_{\bf r r'} = 2\pi f\,.
\end{equation}
Differently from the Lee-Grinstein model \cite{Lee1985}, in Eq. \eqref{eq:2dXYmodel} we consider only junctions with a standard sinusoidal dispersion. In this case the current at each junction is mediated by single Cooper pairs hopping between islands and is defined by: 
\begin{equation} \label{currdef}
I_{\bf r r'} = \frac{2e}{\hbar} E_J \sin \left[\varphi_{\bf r} - \varphi_{\bf r'} -A_{\bf r r'} \right].
\end{equation}

We consider two possible scenarios to numerically estimate the critical current across these arrays. 
The first is the phase-biased scenario \cite{Straley1988} and it reflects experiments in which the system is embedded into an external superconducting loop. 
In this scenario, we need to evaluate the energy of the system with phases $\varphi_{\bf r}$ subject to boundary conditions which are consistent with the controllable external magnetic flux $\Phi_{\rm ext}$. 
In particular, the boundary conditions are imposed by defining two sets of \textit{source} and \textit{drain} islands on the lattice edges, whose phases are constrained such that the phase difference between the two sets matches the external bias $\Phi_{\rm ext}$ (see Appendix \ref{app:rlxsims}). We will refer to all the other superconducting islands as \textit{bulk} islands. We then obtain an estimate of the critical current by maximizing the sum of the currents flowing from the bulk islands to the drain islands as a function of $\Phi_{\rm ext}$ \cite{Teitel1983}.

The second scenario is related instead to current-biased superconducting JJAs. In this case two superconducting leads are connected with the source and drain islands respectively, and we introduce suitable boundary conditions that describe the flow of an external current $I_{\rm ext}$ from each lead to either the source or drain islands \cite{Rzchowski1990}.
In this situation, we check whether there exists a static configuration of the classical phases $\varphi_{\bf{r}}$ that fulfills the current conservation over each bulk superconducting island $\bf{r}$ 
\begin{equation}    \label{currentcons}
\sum_{{\bf r'} \text{n.n.} {\bf r}} \sin \left[\varphi_{\bf r} - \varphi_{\bf r'} -A_{\bf r r'} \right] =0\,,
\end{equation}
where the sum is taken over the nearest neighbor islands ${\bf r'}$ of ${\bf r}$. The critical switching current is then estimated from the largest $I_{\rm ext}$ that allows for a static configuration of the bulk phases.

In both approaches, it is crucial to determine low-energy static configurations of the classical phases as a function of the boundary conditions. In particular, we focus on situations in which either the external flux $\Phi_{\rm ext}$ or the external current $I_{\rm ext}$ are progressively increased in a way mimicking experiments in which these parameters are adiabatically scanned. 
To approximately account for the phase slips occurring in the array during this adiabatic evolution, we resort to simulations of the relaxation dynamics of the superconducting array characterized by local updates of the phases. 
In particular, we discretize the evolution of the boundary conditions in time steps corresponding to small increments of the external flux or current. For each time step, about $10^4$ relaxation moves are performed. Each relaxation move comprises the cyclic update of all the phases of the islands in each of the three sublattices of the dice lattice (see Appendix \ref{app:rlxsims} for more detail). 
These updates locally minimize the energy of the system based on the phases in the other sublattices. Since each time step begins from a configuration corresponding to the previous value of the boundary conditions and returns a new stationary phase configuration, this kind of local update is suitable to mimic realistic phase slips occurring under a slow change of the boundary conditions \cite{Lobb1983}. 
In particular, we observe that the evolution of the stationary configurations of the phases as a function of $\Phi_{\rm ext}$ or $I_{\rm ext}$ typically displays the irregular but local motion of a few vortices in the superconducting array.

\begin{figure}[h!]
  \centering
  \includegraphics[width=\linewidth]{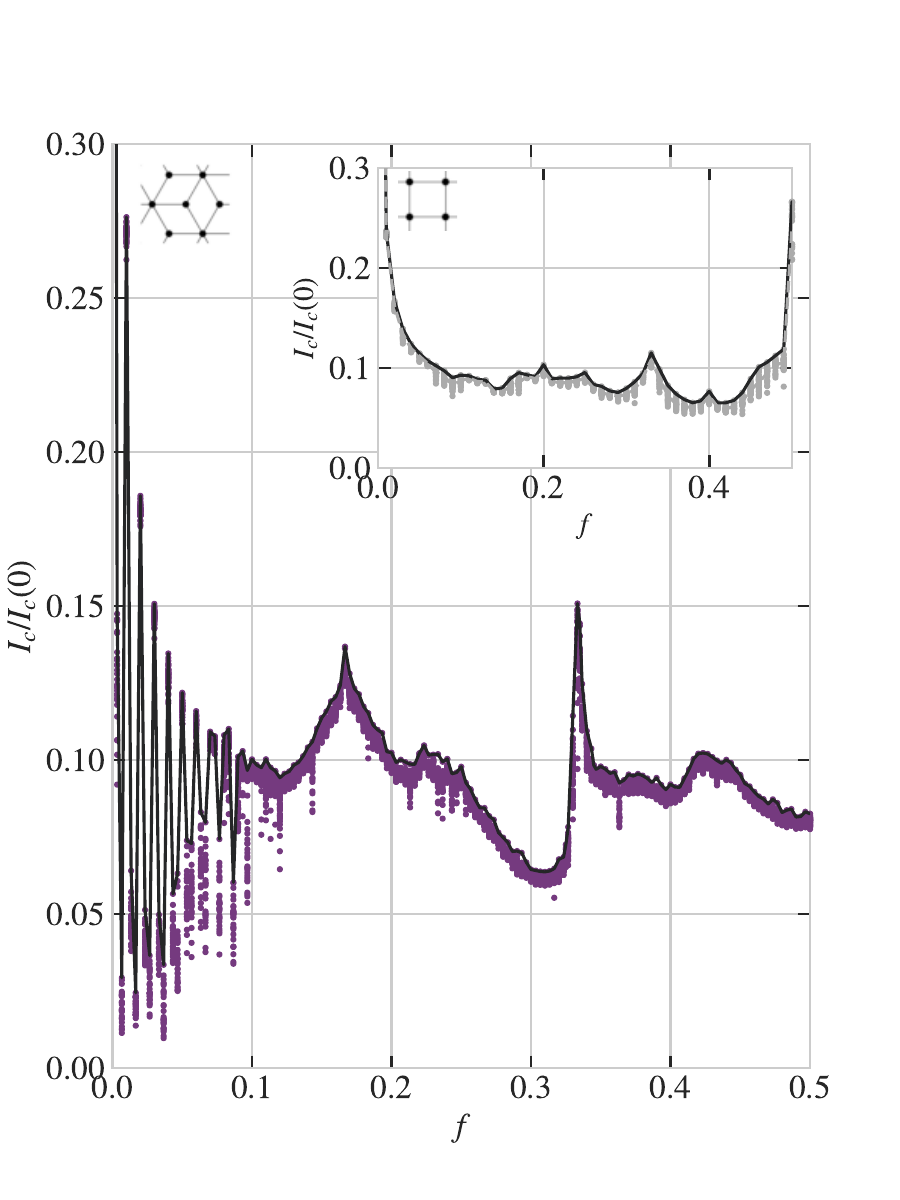}
  \caption{Critical currents as a function of frustration $f$ for a $50 \times 50$ dice lattice calculated using the phase-bias relaxation method. The inset shows corresponding results for a $100 \times 100$ square lattice. The critical currents are normalized by $I_c(0)$. 
  Each purple dot shows the critical current obtained from a single phase configuration. A lower bound on the true critical current can be estimated by taking the maximum of these values (black line). The fringes at low values of $f$ are finite-size effects.
  }
    \label{Fig:criticalcurrent}
  
\end{figure}

We adopted these simulations of the relaxation dynamics to obtain the results on the global critical current displayed in Fig.~\ref{Fig:criticalcurrent}.
The phase-biased and current-biased approaches return analogous results. 
For each value of $f$, we consider the maximal critical current obtained by performing the relaxation dynamics over a state stochastically obtained through simulated Monte Carlo annealing from a high-energy configuration to the desired temperature. In practice, we choose $T = E_J / 100$.
We estimated the critical currents in Fig.~\ref{Fig:criticalcurrent} based on the phase-biased scenario. 
For each value of $f$, the data (purple dots) represent the maximal currents obtained in 100 independent simulations of the relaxation dynamics. For each independent simulation, the maximal current is evaluated after a transient corresponding to the initial increment of the phase bias $\Phi_{\rm ext}$ by a few magnetic fluxes. The black line indicates the highest current obtained for each value of the frustration $f$ and corresponds to a lower bound on the critical current of the system.

The so-obtained profile of the critical current is characterized, first of all, by the global $f=0$ peak corresponding to the absence of magnetic frustration.
In Fig.~\ref{Fig:criticalcurrent}, the large peak at low values of $f$ appears to be modulated by strong oscillations, which, however, constitute a finite-size effect. 
These oscillations have a magnetic flux period  suppressed with the length of the system as $1/L_y$ in the direction $y$ of the current, and are not observed in experimental devices \cite{Bondar2025}, likely due to disorder effects.

By considering larger magnetic fluxes, considerably smaller local peaks of $I_c$ are visible at $f=1/6$ and $f=1/3$. 
These specific frustration values are known to be associated with gapless spectra in the tight-binding approximation of the Hamiltonian \eqref{eq:2dXYmodel} \cite{Vidal1998,Bercioux2024}.
The $f=1/6$ peak appears symmetric with respect to frustration, whereas at $f=1/3$ the peak is clearly asymmetric, a feature consistent with the experimental measurements of the critical current \cite{Bondar2025}. 
Around $f=1/2$, instead, the critical current lies at the center of a valley.
This is expected as Cooper pairs are subject to destructive interference when moving across the dice lattice. This phenomenon is typically called Aharonov-Bohm caging and causes a suppression of the supercurrent carried by Cooper pairs as $2e$-charge carriers become localized \cite{Vidal1998}. This marks the main discrepancy between the dice lattice and other regular arrays: square and triangular lattices, for instance, are characterized by a local maximum of the critical current at $f=1/2$, which, in those cases, corresponds to the onset of regular incompressible vortex lattices \cite{Fazio2001}.

\begin{figure}[t]
  \centering
  \includegraphics[width=\linewidth]{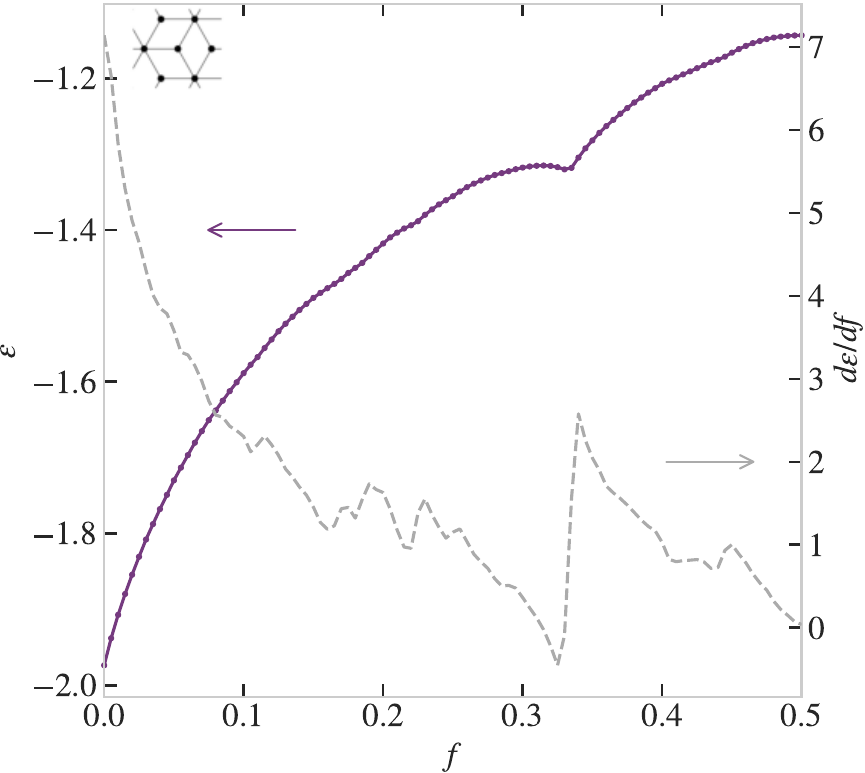}
  \caption{Energy per island (colored curve) and its derivative (grey) vs. $f$ for a $50 \times 50$ dice lattice without source and drain islands, i.e. free boundary conditions, obtained by simulated annealing and a subsequent relaxation step, see App.~\ref{app:rlxsims}.}
  \label{fig:diceenergy}
\end{figure}

A further difference between the behavior of the dice lattice at $f=1/3$ and $f=1/2$ is evident when analyzing the energy density of the system as a function of $f$, cf. Fig.~\ref{fig:diceenergy}. 
Analogously to the square lattice (see Ref. \cite{Lankhorst2018}) the critical current peak at $f=1/3$ is matched by a local minimum of the energy density $\varepsilon$ in the shape of a cusp with a discontinuous derivative $d\varepsilon/df$. On the contrary, the full frustration reached at $f=1/2$ yields the global maximum of the energy density $\varepsilon$.
We observe no clear local minimum in the energy corresponding to the peak in the critical current at $f = 1/6$.

The behavior we derived from the XY model in Eq.~\eqref{eq:2dXYmodel} qualitatively captures all the main features in the experimental data. The main discrepancy concerns the global minimum of $I_c$: in the numerical simulations, the minimum of the critical current lies just before the $f=1/3$ peak, whereas in the experimental data it occurs at $f=1/2$. Additional discrepancies involve the relative heights of the current peaks: In the simulations, the peak at $f=0$ is overestimated, and the peaks at $f=1/3$ and $f=1/6$ have comparable heights; this is in contrast with the experimental data where the peak at $f=1/3$ is markedly stronger than the one at $f=1/6$ \cite{Bondar2025}. The differences between numerical estimates and experimental data on the critical currents can be attributed to the effects of disorder in the system (see Sec. \ref{sec:dis}).

Experimentally, the behavior of the critical temperature displays a peak structure that qualitatively matches $I_c$ \cite{Abilio1999,Tesei2006}.

\subsection{Harmonic analysis of bulk supercurrents}
\label{sec:harmonic_decomp_of_supercurrent}

The XY model in Eq. \eqref{eq:2dXYmodel} describes an array of Josephson junctions with a simple sinusoidal energy-phase relation. For hybrid semiconductor-superconductor arrays, this corresponds to experimental regimes of low transparency of the semiconductor channels mediating the Cooper pair transport between neighboring superconducting islands \cite{Beenakker1991,Kringhoj2018}.

Despite the resulting sinusoidal current-phase relation \eqref{currdef} between neighboring islands,  many-body and interference effects may favor the onset of supercurrents between arbitrary source and drain islands that display higher harmonics contribution as a function of the corresponding gauge-invariant phase differences.

Therefore, a general supercurrent between source and drain islands can be described using a current-phase relation of the kind:
\begin{equation} \label{CPR}
I(\delta \theta) = -\frac{2e}{\hbar}\frac{d}{d\delta \theta} E(\delta \theta) = \frac{2e}{\hbar}\sum_n I_n \sin \left(n\delta \theta\right)\,,
\end{equation}
where 
$\delta \theta$
represents the gauge-invariant phase difference between source and drain, and $n$ labels the related current harmonic.

The index $n$ is directly related to the charge of the current carriers. This can be verified by considering an ideal setup in which a phase bias $\Phi_{\rm ext}$ is imposed exclusively between two chosen superconducting islands in the array. Such setup would correspond to an experiment in which a superconducting bridge connects the two islands and the magnetic flux $\Phi_{\rm ext}$ pierces the superconducting loop formed by the bridge and the Josephson junction array. In this case, 
$\delta \theta = 2e \Phi_{\rm ext}/\hbar$
, and each harmonic $I_n$ represents a current contribution with the flux periodicity of 
$\sin \left(n\delta \theta\right)$
. This periodicity is dictated by the Aharonov-Bohm phase $2n e \Phi_{\rm ext} / \hbar$ which carriers with charge $2ne$ acquire while undergoing the interferometric path. Therefore, the supercurrents flowing through interferometric loops connecting pairs of islands of the arrays provide direct information about the charge of the carriers.

The picture described by Eq. \eqref{CPR} is supposed to hold when considering the statistical average of the energy 
$E(\delta \theta)$
over all possible phase configurations at a given temperature. 
However, when considering the current through the entire lattice as we did in Sec.~\ref{sec:criticalcurrent}, the evolution of a specific vortex configuration under the adiabatic change of the phase-bias parameter $\Phi_{\rm ext}$ results in a behavior of $I(\Phi_{\rm ext})$ typically affected by discontinuities. These jumps are caused by random phase slips that correspond to the stochastic motion of magnetic vortices in the array caused by thermal fluctuations.

On the dice lattice, the effect of these random phase slips is particularly disruptive for values of the frustration parameter $f$ such that there is an extensive degeneracy of vortex configurations with minimal energy. These extensive degeneracies occur, in particular, at $f=1/3$ and $f=1/2$ \cite{Korshunov2005,Korshunov2005b} and result in non-periodic patterns $I(\Phi_{\rm ext})$ of the current flowing across the whole array as a function of the phase bias $\Phi_{\rm ext}$. Indeed, the analysis of the vortex configurations during the simulations of the relaxation dynamics in the phase bias scenario reveals irregular transitions between degenerate metastable vortex configurations that usually differ locally just by the position of a few vortices, but are characterized by different global currents. For $f=1/3$ and $f=1/2$ these irregular phase slips prevent the general observation of a periodic $I(\Phi_{\rm ext})$, such that the analysis of the global current as a function of the phase bias is impractical to extract information about the transport contribution of Cooper pairs and quartets.

\begin{figure}[t!]
  \centering
  \includegraphics[width=\linewidth]{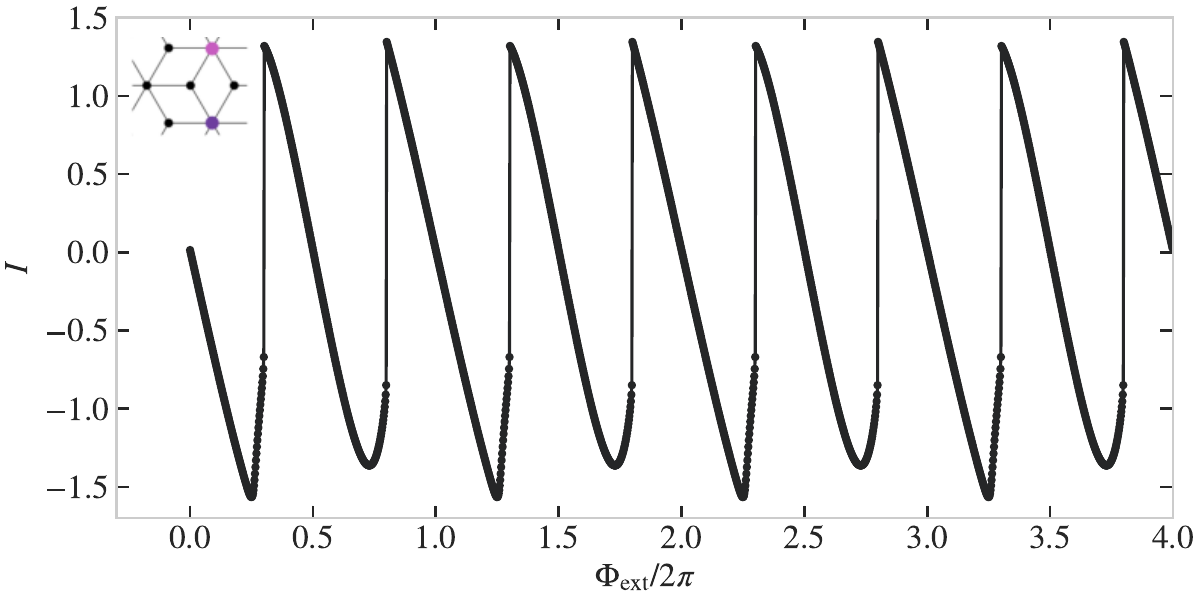}
  \caption{
  Typical current-phase relation for $f=1/3$ for a phase-biased supercurrent between source (purple) and drain (pink) islands situated in the middle of a $50 \times 50$ dice lattice. $I$ is expressed in units of $\frac{2e}{\hbar}E_J$ and $\delta \Phi_{\rm ext}=2\pi/500$.}
  \label{Fig:currentphase}
\end{figure}

The lack of a periodic current-phase relation is mitigated when considering a pair of source and drain islands at short distance (up to a few unit cells). In this scenario, the phase slips as a function of $\Phi_{\rm ext}$ correspond to the regular motion of vortices around either of these islands (cf. Sec.~\ref{sec:vort_dyn}). Based on our simulation of the relaxation dynamics, we observe that such motion typically involves only a small subset of plaquettes around source and drain islands and, after a possible short transient that follows the initial quenching protocol, it repeats periodically with $\Phi_{\rm ext}$, see Fig.~\ref{Fig:currentphase}.

\begin{figure}[h!]
  \centering
  \includegraphics[width=\linewidth]{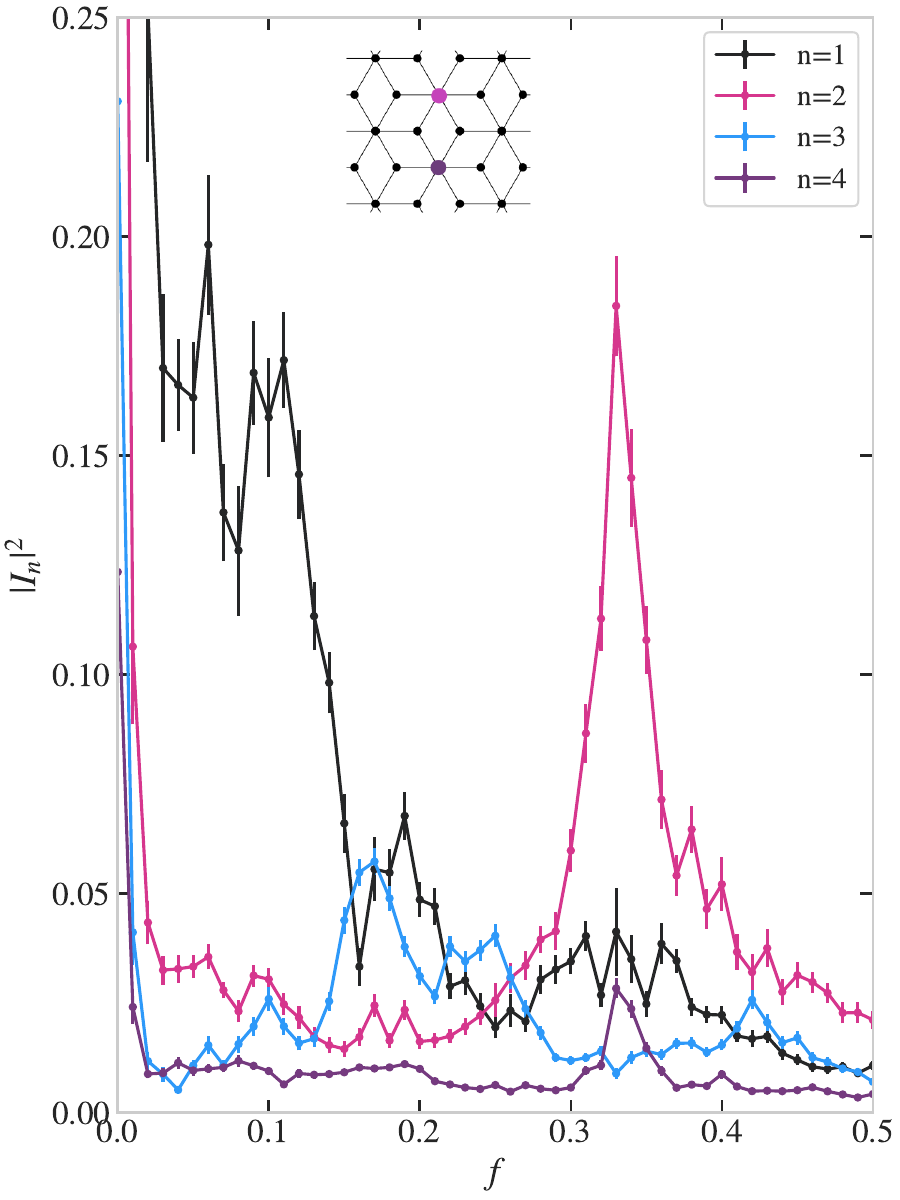}
  \caption{Harmonics of the supercurrent across a single rhombus as a function of $f$.
  The source and drain islands for this supercurrent are highlighted in purple (source) and pink (drain) in the inset. These islands are situated in the middle of a $50 \times 50$ dice lattice, and the supercurrent is determined by the introduction of a phase bias between them, see App.~\ref{app:rlxsims}. 
  Points and error bars indicate the mean and standard error of the mean from 100 independent simulations for each value of $f$. }
  \label{Fig:Fourier}
\end{figure}

For the case of source and drain islands with connectivity six (usually called hubs) taken at opposite positions in the same rhombus, the first Fourier components of $I(\Phi_{\rm ext})$ are illustrated in Fig.~\ref{Fig:Fourier}.
Here and in the following, the components $I_n$ are expressed in units of $\frac{2e}{\hbar}E_J$ (see App. \ref{app:rlxsims} for more details).
The $f=0$ peak, as expected, is predominantly determined by the $n=1$ harmonic. This corresponds to the $2e$ superconducting phase that characterizes the dice lattice at small frustration. 

The peak at $f=1/6$, instead, displays the largest value of the $I_3$ component, corresponding to a supercurrent mostly mediated by 6e Cooper multiplets. This result is counterintuitive since 6e carriers are subject to destructive interference across a single rhombus and it emphasizes the fact that the current across a single plaquette is determined by the phase configuration of the whole lattice, rather than the single rhombus behavior.

Crucially, at $f=1/3$ we observe a broad peak of the current across the rhombus, caused by a sharp increase of the $I_2$ component which reaches its global maximum. Around this peak, the $I_2$ current amplitude dominates all the other components, and it reaches a value approximately five times larger than $I_1$. 
This suggests that indeed the supercurrent transport is predominantly mediated by Cooper quartets for frustration values around $f = 1/3$ and it corroborates Korshunov's picture \cite{Korshunov2005b}: the low-energy phase of the dice lattice at $f=1/3$ appears to be characterized in terms of disordered superconducting phases $\ee^{i\varphi}$, due to the proliferation of domain walls between several vortex patterns, but quasi ordered $\ee^{i2\varphi}$ components, consistently with a $4e$ phase with confined half-vortices, cf. Sec.~\ref{sec:emergent_4e}.  

The situation is considerably different at $f=1/2$, where all the harmonic components of the current are considerably suppressed, with the $4e$ contribution still being the largest one. The resulting current across a rhombus displays a local minimum analogously to the critical current of the full system.

So far, we considered the current across a single rhombus. A similar behavior of the current harmonics for $f=1/3$, however, is displayed when considering the transport across arbitrary pairs of islands with sixfold connectivity.

\begin{figure}[t!]
  \centering
  \includegraphics[width=\linewidth]{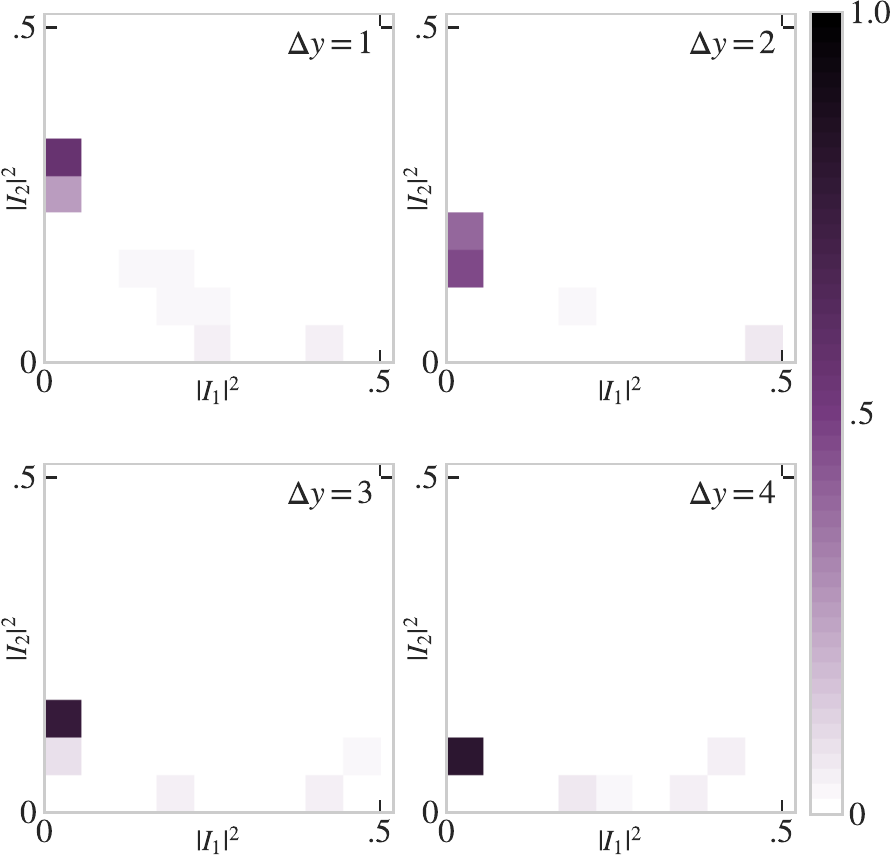}
  \caption{Two-dimensional histograms showing the joint probability of $(|I_1|^2,|I_2|^2)$ at $f=1/3$ for different vertical distances $\Delta y$ (in units of $\sqrt{3}a$) between source and drain located on sixfold coordinated islands near the middle of a $50 \times 50$ dice lattice. Each histogram is based on 100 independent runs, and the color indicate the fraction of runs within a box.}
  \label{Fig:harmonicshistograms}
\end{figure}

Fig.~\ref{Fig:harmonicshistograms} displays the joint distribution of $I_1$ and $I_2$ obtained by sets of 100 independent stochastic annealing initializations of the system at temperature $T = E_J/100$ across multiple rhombi along the $y$ direction. The distances $\Delta y$ between source and drain islands are expressed in units of $\sqrt{3}a$, where $a$ is the lattice spacing. The second harmonic clearly dominates at the considered distances, despite the presence of rare configurations with considerably higher $I_1$ components. A similar behavior is observed for separations of the source and drain island along the inequivalent $x$ direction (see Appendix \ref{app:additionaldata}). 
This general dominance of the second harmonic at $f=1/3$, however, is a characteristic of the sublattice of the islands with sixfold coordination. The currents calculated by biasing islands in the other sublattices, instead, display a standard behavior with $I_1 \gg I_2$.

The analysis of the Fourier components of the supercurrents between pairs of sixfold-connected islands in the bulk of the array is not sufficient to rigorously assess the nature of the superconducting phase in this regime. However, the dominance of the $I_2$ component in the broad range from $f=1/3$ to $f=1/2$ (and, symmetrically, to $f=2/3$) suggests the possibility that there exists an extended $4e$ phase in this frustration interval, which displays the maxima of its critical current and critical temperature at $f=1/3$. The detection of possible phase transitions between different superconducting phases as a function of $f$ is a challenging problem to be numerically addressed, as fine variations of the parameter $f$ result in large magnetic unit cells, thus in difficulties to access the thermodynamic limit. We leave this analysis for future works. 
In the following section, we will focus instead on the specific $f=1/3$ case.

\subsection{Effects of disorder and robustness} \label{sec:dis}

\begin{figure}[h!]
  \centering
  \includegraphics[width=1\linewidth]{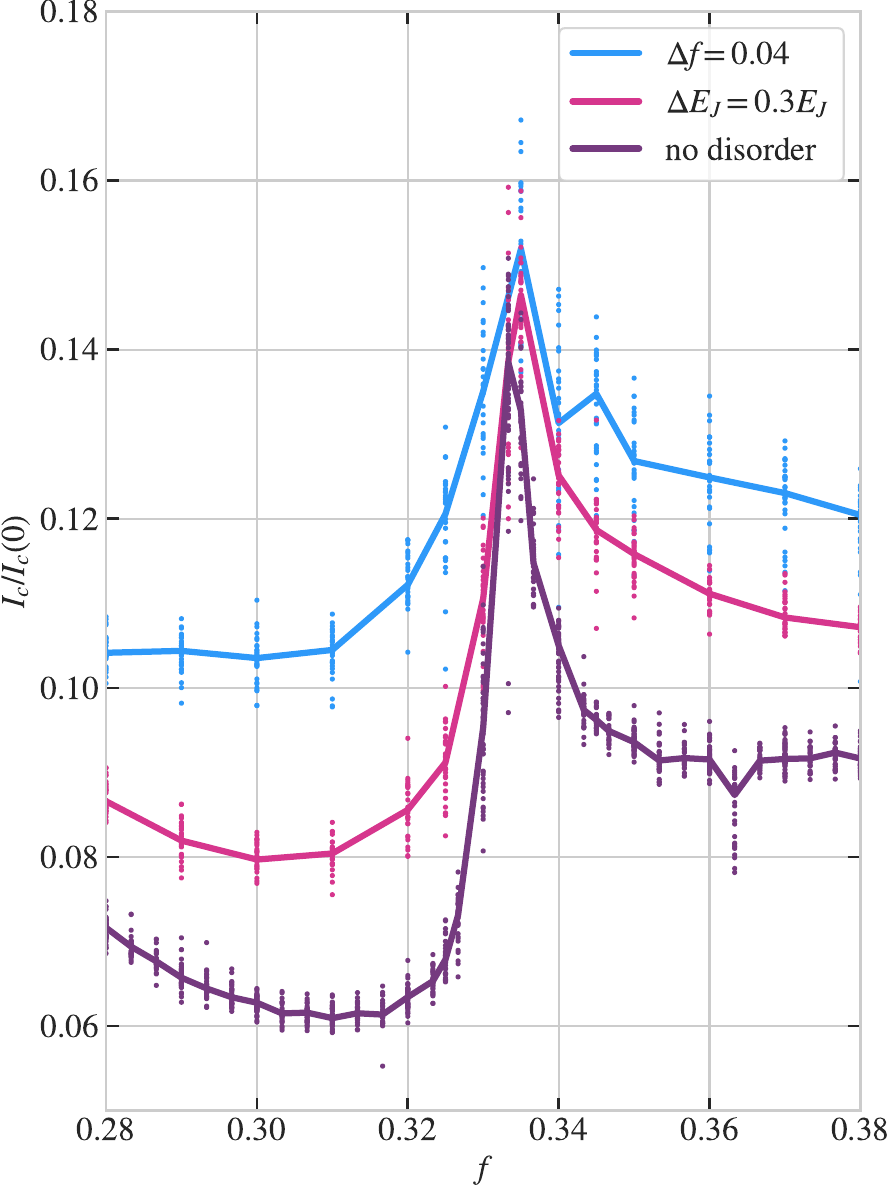}
  \caption{Critical currents around $f=1/3$ for two disorder scenarios: Flux disorder with $\Delta f=0.04$ (blue), and Josephson disorder with $\Delta E_J = 0.3 E_J$ (pink). Critical currents with no disorder are also shown (purple). The curves indicate the median of critical currents obtained in 30 independent runs (dots). The critical currents are normalized to the $f=0$ value of the critical current in the case of no disorder. 
  \label{Fig:disorder_criticalcurrents}}
\end{figure}

Disorder can play an important role in the physics of JJAs. 
To investigate the stability of the features discussed above towards disorder, we repeat the critical current calculations performed in Sec.~\ref{sec:criticalcurrent} including two kinds of disorder:   (i) disorder $\Delta f$ in the flux per plaquette, caused by both fabrication details that affect their area, and a possible position dependence of the magnetic field; (ii) disorder $\Delta E_J$ in the Josephson energies $E_J$, determined by both the fabrication of the junctions and the disordered potential landscape of the semiconductor environment.

In Fig.~\ref{Fig:disorder_criticalcurrents}, we show that the critical current peak at $f=1/3$ is rather robust against disorder. We display, in particular, the critical current around $f=1/3$ for either flux disorder (blue) or Josephson disorder (magenta). 

In the case of $\Delta f$, the disorder is numerically implemented by adding a gauge potential disorder to all the phases $A_{\bf{rr'}}$ in Eq. \eqref{eq:2dXYmodel}. These fluctuations are chosen to be Gaussian with standard deviation $2\pi \Delta f/2$, with $\Delta f =0.04$.

In the case of $\Delta E_J$, each cosine term in the Hamiltonian is multiplied by a random Josephson energy selected as a Gaussian random number with mean $E_J$ and standard deviation $\Delta E_J=0.3E_J$.  

In general, we observe that the critical current is very robust against Josephson disorder irrespective of $f$. The $30\%$ disorder showed in Fig.~\ref{Fig:disorder_criticalcurrents} is indeed considerably stronger than the expected experimental scenario. Choosing a more realistic deviation of only $0.1 E_J$ (not shown) results in a critical current pattern that is very similar to the situation with no disorder around both $f=0$ and $f=1/3$. The robustness of the critical current against $\Delta E_J$ can also be appreciated by observing the effect of disorder on the $f=0$ peak (not shown) which, for $\Delta E_J =0.3 E_J$ is slightly suppressed to approximately $0.88 I_c(0)$.

In comparison, a small value $\Delta f =0.04$ of flux disorder dampens the $f=0$ peak down to $0.94 I_c(0)$, and, in general, we observe a stronger influence of the magnetic flux fluctuations $\Delta f$ over the critical current profile with respect to Josephson energy fluctuations $\Delta E_J / E_J$.

The peak at $f=1/3$ is resilient against flux disorder for $\Delta f \lesssim 0.05$, and we observe that the introduction of a moderate $\Delta f$ actually increases the critical current in the whole region around $f=1/3$. We notice, in particular, that both Josephson and flux disorder contribute to lift the minimum of the critical current in the clean XY model (appearing around $f\sim 0.31$; see also Fig.~\ref{Fig:criticalcurrent}). Moderate flux disorder seems indeed responsible for the disappearance of this global minimum, such that, in disordered arrays, the minimum of the critical current is reached at $f=1/2$ instead. Finally, we also observe that for both the introduction of $\Delta E_J$ and $\Delta f$ the typical asymmetry of the $f=1/3$ peak is preserved. These features suggest that the experimental platforms analyzed in Ref. \cite{Bondar2025} are indeed affected by flux disorder.

\section{Emergent $4e$ phase at $f=1/3$} \label{sec:13}
\label{sec:emergent_4e}
In the previous section we observed the onset of a critical current peak at $f=1/3$, we observed its stability under flux and Josephson energy disorder and we showed that such peak coincides with a dominant contribution of the transport or Cooper quartets to the supercurrents in the bulk of the Josephson junction dice lattice. In the following we explore additional signatures of the emergent $4e$ phase at $f=1/3.$

\subsection{Ground state properties and half-vortices}
\label{sec:Ground state properties and half-vortices}
At vanishing temperature, the ground state vortex configurations of the dice lattice XY model at frustration $f=1/3$ can be mapped to the ground state configurations of the classical antiferromagnetic Ising model on a triangular lattice \cite{Korshunov2005b}.
Each hexagon is indeed decomposed into three rhombi, and each of them may or may not host a magnetic vortex. These two possibilities correspond respectively to the ferromagnetic or antiferromagnetic link configurations of the Ising model defined on the triangular sublattice of the superconducting islands with coordination six. 
The ground state configurations consist of patterns of vortices, in which each hexagon hosts exactly one vortex. This reflects the fact that each hexagon is characterized by a $\Phi_0$ flux for $f=1/3$, such that, at low energy, one expects each hexagon to display a unitary vorticity. For each hexagon, the three possible minimal energy choices display a perfect degeneracy, and, at low energies, two vortices never occupy adjacent rhombic plaquettes.

Therefore, an extensive degeneracy of ground states analogous to the antiferromagnetic Ising model on the triangular lattice was predicted for the XY model on the dice lattice at $f=1/3$~\cite{Korshunov2005b}. This entropy at zero temperature corresponds to  $S_\mathrm{I} = \tfrac{3}{\pi} \int_0^{\pi/6} \ln \left( 2 \cos \omega\right) \, \mathrm{d}\omega \simeq 0.323$ per unit cell \cite{Wannier,Wanniererratum}. 

\begin{figure}
    \centering
    \includegraphics[width=0.95\linewidth]{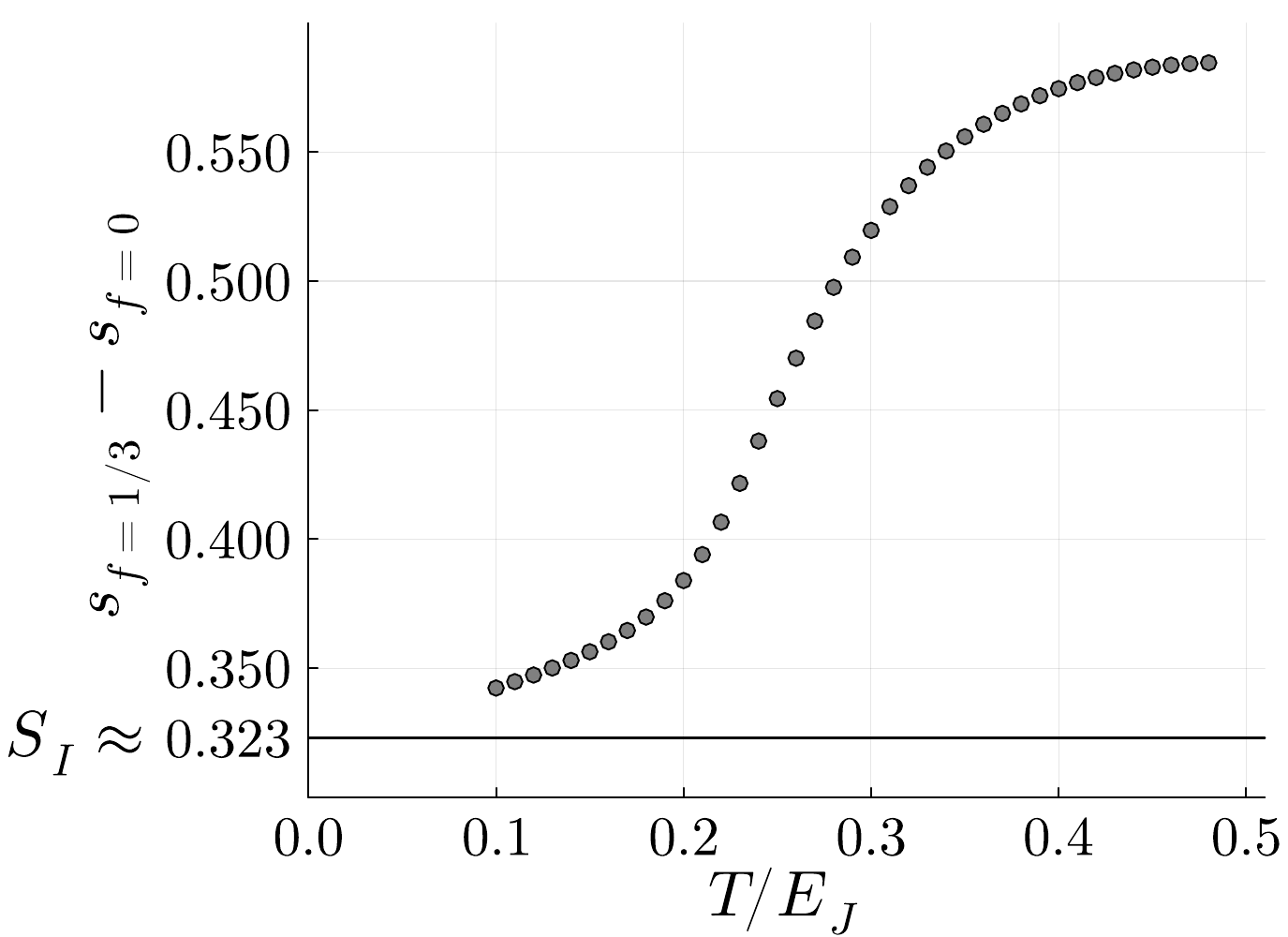}
    \caption{Entropy density difference of the XY-model at frustration $f=\frac{1}{3}$ and $f = 0$. We find that the difference approaches the residual zero-point entropy of the triangular lattice antiferromagnetic Ising model (horizontal line). At temperatures below $T/E_J = 0.1$, the tensor-network calculations did not converge consistently, and data are not reported. }
    \label{fig:entropyplot}
\end{figure}

We can numerically isolate this discrete Ising-like contribution to the entropy density, stemming from the degeneracy of different vortex configurations. By using a tensor network representation of the partition function of the XY model \cite{Vanderstraeten2019_BKT_TN} on the dice lattice at $f=1/3$, specifically the one described in \cite{Song2023_general_theory_XY}, we can compute the entropy density. 
However, in addition to the discrete Ising-like entropy contribution of the vortex configurations, there is an entropy offset associated with the continuous nature of the superconducting phases characterizing the XY model. 
To remove this continuous offset, we show in Fig.~\ref{fig:entropyplot} the difference of the entropy density of the XY model at $f=\frac{1}{3}$ and at $f=0$.
As the unfrustrated model at $f=0$ has no discrete contribution to its entropy density, but is characterized by the same continuous variables, we expect their difference at $T=0$ to match the entropy density the antiferromagnetic Ising model on the triangular lattice at low temperature. In Fig.~\ref{fig:entropyplot}, we show indeed that this difference seems to converge towards the predicted value  $S_\mathrm{I}$ for small temperatures (horizontal line). 
Let us note that for very low temperatures ($T\lesssim 0.1J$), we have experienced convergence problems for the adopted TN-algorithms \cite{nishino1996corner_CTMRG1, nishino1997corner_CTMRG2, Corboz2011_CTMRG4.1_unitcell, corboz2014competing_CTMRG5_projectors, Schmoll2021_Heisenberg} and the data are not displayed. This behaviour has also been observed in previous works on two-dimensional tensor networks \cite{Vasseur2024, Schuch2025, fPEPS_ours} and occurs for other values of the frustration $f$ as well. 

\begin{figure}[t]
\includegraphics[width=\columnwidth]{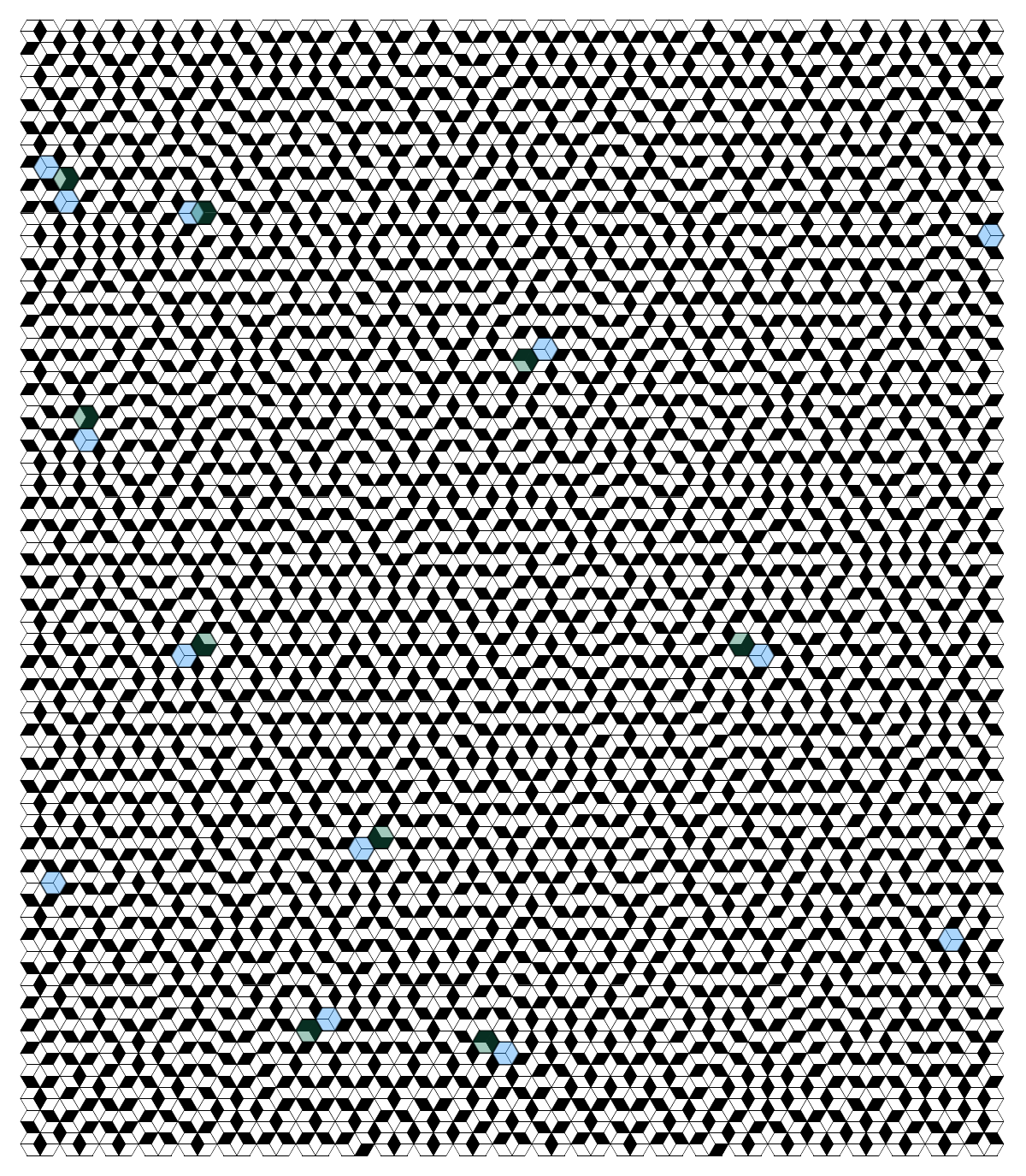}
\caption{Example of a vortex configuration with rare occurences of half vortices. The blue and green hexagons respectively show the half-vortex and half-antivortex excitations. In the interior only bound pairs of half-vortices exits, while close to the vertical edges there are some single half-vortices. The configuration shown was taken at $T=0.187E_J$ with open boundary conditions at the vertical edges and constrained phases on the horizontal ones, cf. App.~\ref{app:rlxsims}.}
\label{fig:vortexconfig}
\end{figure}

The extensive entropy can also be seen as the contribution from zero-energy domain walls between families of degenerate periodic ground-state vortex patterns \cite{Korshunov2005b}. In the low-temperature phase, these domain walls are either closed or originate from boundaries and excitations over the ground state. Since there is no tension associated to them, they cover the full array in a dense way and the low-temperature vortex configurations are thus disordered (see Fig. \ref{fig:vortexconfig}).

The lowest-energy excitations of the system, instead, are given by hexagons with either two vortices or no vortex at all.
These low-energy excitations correspond to the joining of three such domain walls and constitute the half-vortices that characterize the low-temperature phase of the dice lattice \cite{Korshunov2005b}.

\begin{figure}[t]
\includegraphics[width=\columnwidth]{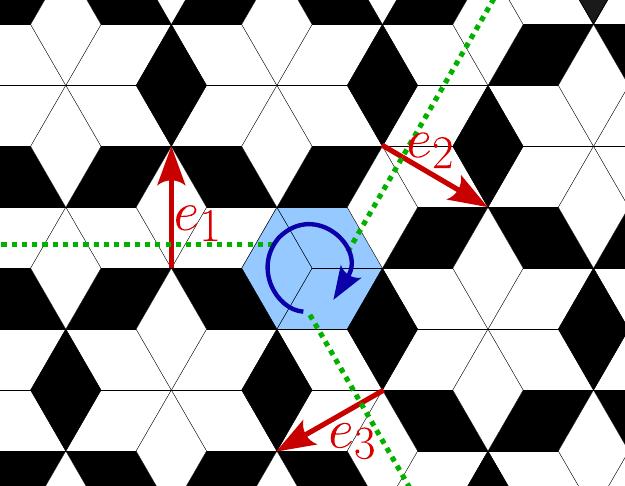}

\caption{Example of half-vortex for $f=1/3$: Black and white rhombi respectively depict rhombi with and without vortices. In this example, the half vortex corresponds to an excitation (light blue hexagon without vortices) at the joining of three domain walls (green dotted lines corresponding to stripes of white rhombi) between three possible ground state vortex configurations. The domain walls constitutes defect lines associated to translations by the vectors ${\bf e_a}$ (red) of the vortex patterns of the ground states. Each domain wall also corresponds to a shift by $\pi$ of the superconducting phases in one of the dice sublattice \cite{Korshunov2005b}. Moving along the blue path around the excitation yields an overall discontinuity by $\pi$ of all phases, as expected for a half-vortex. 
}
\label{fig:vortex}

\end{figure}
This is illustrated in Fig.~\ref{fig:vortex}, where we depict three domain walls (green dotted lines) between different honeycomb-like ground state patterns of the vortices (black rhombi). Black and white rhombi respectively represent plaquettes $\mathcal{P}$ with phase vorticity 
\begin{equation}
    v = \frac{1}{2\pi}\sum_{\langle {\bf r},{\bf r'}\rangle \in \mathcal{P}}\left(\varphi_{\bf r} - \varphi_{\bf r'}\right),
\end{equation}
taking values 1 and 0.
The represented honeycomb domains are a subset of all possible ground state patterns and the half-vortex in Fig.~\ref{fig:vortex} constitutes the simplest instance of three domains joining, as all the involved vortex patterns are equivalent to each other through translation.

In Fig.~\ref{fig:vortex}, each domain wall corresponds to a stripe of rhombi without vortices (white rhombi along the green dotted lines) and can be interpreted in two complementary ways: (i) it corresponds to a line defect associated to the translation of the ground state vortex pattern by a lattice vector ${\bf e_j}$; (ii) it corresponds to a shift by $\pi$ of the superconducting phases $\varphi_{\bf r_j}$ of the sublattice $j$. Therefore, when considering a path that surrounds the central excitations, we observe that the vortex lattice is translated by the sum of the three lattice vectors ${\bf e_1} +{\bf e_2} + {\bf e_3}= {\bf 0}$ such that, overall, it goes back to itself. The superconducting phases, however, present discontinuities that correspond to an overall $\pi$ phase shift in all the sublattices. Such discontinuity is reminiscent of a defect line in the Lee-Grinstein model \cite{Lee1985}, but, in the dice lattice, it is split into three different defect lines corresponding to each of the sublattices. Therefore, in the dice lattice at $f=1/3$, the half-vortices are excitations appearing at the joining of three zero-energy domain walls rather than at the ending point of a single line.

We emphasize that configurations with single isolated half-vortices, as the one depicted in Fig.~\ref{fig:vortex}, are not typical: in the low-temperature phase, half-vortices are confined and appear in isolated pairs composed by a half-vortex and a half-antivortex (see Fig.~\ref{fig:vortexconfig} and the central column in Fig.~\ref{Fig:Mitsubishi}); in the high-temperature phase, instead, despite being deconfined, their density is large and it is not common to find them isolated.

In general, also the domain walls between vortex patterns that are not equivalent through translation generate a half-vortex excitation at their joining. Additionally, half-antivortices, characterized by two neighboring plaquettes with vorticity 1, share an analogous behavior with half-vortices and appear at the joining of triplets of domain walls. 
A comprehensive analysis of all possible domain walls can be performed based on the mapping from the ground state domains onto a solid-on-solid model \cite{Korshunov2005b}: Korshunov showed that the half-vortices are mapped in this way into topological excitations corresponding to screw dislocations with Burger numbers $\pm 3$ (concerning the relationship between vortices in the XY model and dislocations of the solid-on-solid model see, for instance, Ref. \cite{vaneerden1978}).

\subsection{Vortex dynamics in the superconducting bridge construction}
\label{sec:vort_dyn}

\begin{figure}[t!]
  \centering
  \includegraphics[width=1.\linewidth]{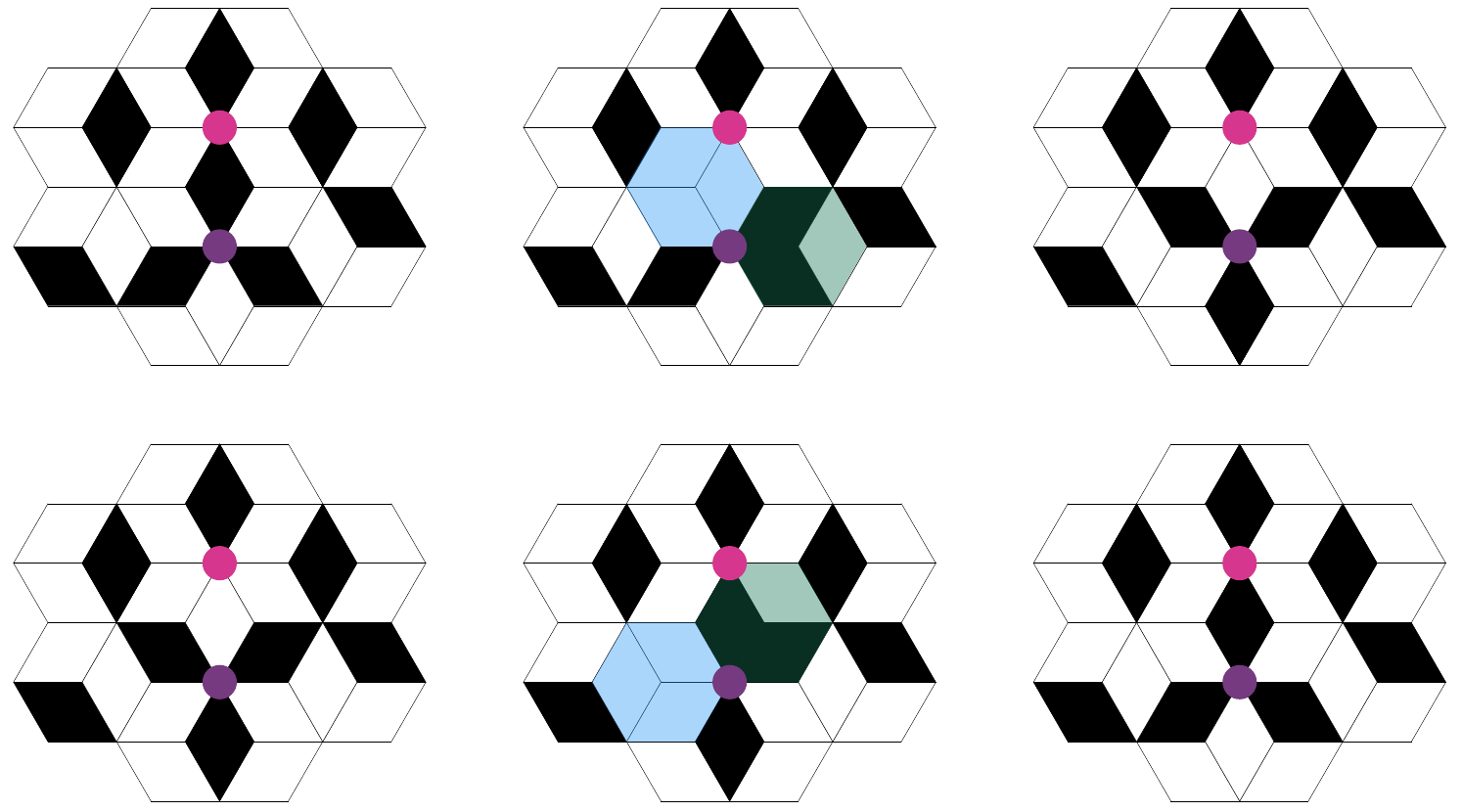}

    \caption{Vortex patterns near the source (purple) and drain (pink) islands obtained for $f=1/3$ in the relaxation simulation corresponding to the current shown in Fig.~\ref{Fig:currentphase}. The upper panel shows the vortices at $\Phi_{\rm ext}/2\pi = \{ 3.282,3.284,3.304 \}$ (left to right) and the lower panel at $\Phi_{\rm ext}/2\pi = \{ 3.784,3.786,3.802 \}$. A black (white) rhombus indicates a vorticity $1$ ($0$).
  The blue and green hexagons respectively show the half-vortex and half-antivortex excitations.}
  \label{Fig:Mitsubishi}
\end{figure}

The presence of half-vortices and their connection to domain walls is intimately linked to the large second harmonic contribution of the current-phase relation between pairs of islands in the sublattice with sixfold coordination at $f=1/3$, cf. Fig.~\ref{Fig:Fourier} and Fig.~\ref{Fig:harmonicshistograms}. 
The corresponding halved periodicity of the (two-island) current-phase relation visible in Fig.~\ref{Fig:currentphase} is related to the periodic formation and annihilation of pairs composed of a half-vortex and a half-antivortex. 
This is illustrated in Fig.~\ref{Fig:Mitsubishi}, where we show the vortex dynamics corresponding to the adiabatic evolution of the two-island current. The six panels depict a typical evolution of the vortex patterns around the source and drain islands during a (double) period corresponding to the increment of $\Phi_{\rm ext}$ by $2\pi$.

During most of the cycle, there are no excitations in the system, which lies in a ground state vortex configuration, as the ones illustrated in the left and right columns of Fig.~\ref{Fig:Mitsubishi}. The top left panel depicts the state of the system for $\Phi_{\rm ext}=0$. This configuration remains stable up to a critical flux-bias value ($\Phi_{\rm ext}\sim 0.28 \times 2\pi$ in the example).
When the external bias reaches this critical value, an excitation pair constituted by a half-vortex and a half-antivortex is nucleated in two neighboring hexagons, stemming from either the source or the drain island. In Fig.~\ref{Fig:Mitsubishi}, this corresponds to a phase slip in which one of the vortices (black rhombi) changes its position, moving counterclockwise by one rhombus around the drain island. 
By increasing the external flux bias further, other vortices rotate around the drain island, in a counterclockwise motion that corresponds to the winding of the excitation pair around the drain and their subsequent annihilation.
In Fig.~\ref{Fig:Mitsubishi} we observe that, indeed, when $\Phi_{\rm ext}$ is increased beyond a second threshold ($\sim 0.30 \times 2\pi$ in this example), the system is found again in a ground state. 
However, the three black rhombi surrounding the drain island have rotated by $2\pi/3$ with respect to the initial configuration. Such rotation can indeed be expected when the phase of the drain island is shifted by $\pi$, and it corresponds to the fact that the vortex-antivortex pair surrounded the drain island before reannihilating, thus effectively creating a minimal instance of a  Lee-Grinstein defect line enclosing the drain, consistently with its $\pi$ phase shift. 
The panels in the top and bottom row of Fig.~\ref{Fig:Mitsubishi} are related by displacements of $\Phi_{\rm ext}$ by half a period. We observe that in the second half of the period, the same dynamics repeat: a pair of half-vortex and half-antivortex excitations is created, winds for the second time around the drain island, and annihilates. After this second winding, the ground state is restored to its original configuration.

In general, an effective current-phase relation with a dominant second harmonic corresponds to two cycles in which an excitation pair is created, winds around either the source or the drain island, and reannihilates. 
Consequently, the ground state configuration of the vortices cycles between two configurations, connected by excited states that display a pair of half-vortices.
Interestingly, we observe that the outlier instances characterized instead by $I_1>I_2$ in the statistical distribution of the harmonics of the current-phase relation depicted in Fig.~\ref{Fig:harmonicshistograms} correspond to qualitatively different kinds of dynamics, in which multiple pairs of excitations are created and get reannihilated only after a full period in $\Phi_{\rm ext}$, instead of half of a period.

\subsection{Signatures of the $4e$ phase and the BKT phase transition} \label{sec:4ecorr}

The analysis of the low-temperature regime of the dice lattice JJAs demand a careful characterization of the possible superfluid phases originated by different charge carriers.
In general, we may expect three possible phases: (i) the standard superconducting phase characterized by a condensation of Cooper pairs and the confinement of both integer and fractional magnetic vortices; (ii) the $4e$ superfluid phase characterized by a condensation of Cooper quartets, half-vortices confined in pairs, and the proliferation of $\pi$-domain walls of the superconducting phase; (iii) the normal insulating phase without any superconducting order parameter, obtained from the deconfinement of either integer or fractional vortices.

Within this framework, the onset of half-vortex excitations and the dominant second harmonic of the supercurrents suggest that the low-temperature superconducting phase observed at $f=1/3$ is mediated by Cooper quartets and provides an example of the $4e$ phase.
However, our previous analysis of the two-island supercurrents and the related system evolution under a slow increase of the phase bias are limited to relatively small distances between source and drain. They are thus insufficient to rigorously assess whether the superconducting regime at $f=1/3$ truly corresponds to the $4e$-superfluid phase.

To corroborate this hypothesis, we analyze two main characteristics of the low-temperature regime: the superconductor-insulator phase transition as a function of temperature and its correlation functions.

\begin{figure}[t!]
  \centering
  \includegraphics[width=1\linewidth]{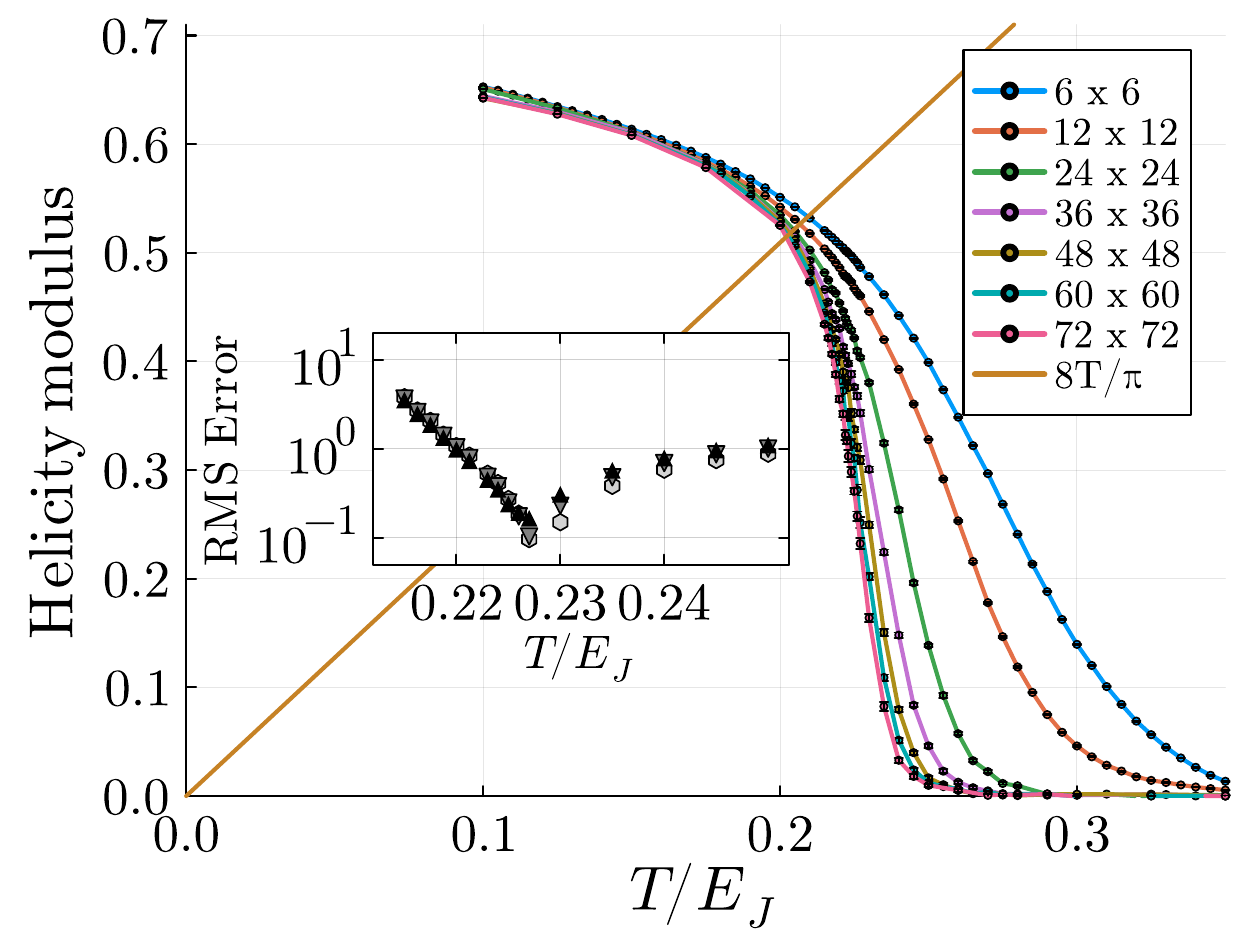}
  \caption{Helicity modulus for phase twists along the y-direction vs. $T/E_J$ for $f=1/3$. Each curve is obtained by Monte Carlo simulations on an $L \times L$-lattice, with periodic boundary conditions.  The size $L$ is specified in the legend. The line has slope $8/\pi$. \textit{Inset:} Root mean-squared error of the fit of the finite-size corrections of the helicity data in Eq. \eqref{Weber} for every fixed temperature \cite{WeberMinnhagen}. We find the best fit at $T=0.227\pm 0.003 E_J$, indicating the expected criticality. The light-grey hexagons are based on a fit including data of sizes 6 to 48, the grey triangles include data of sizes 6 to 60 and the black up-triangles include data of sizes 6 to 72.}
  \label{fig:dicehelicity}
\end{figure}

Fig.~\ref{fig:dicehelicity} shows the helicity modulus of the XY model at $f=1/3$ for various system sizes obtained from Monte-Carlo simulations. For a BKT transition  due to the unbinding of half-vortices, renormalization group arguments predict that the value of the helicity modulus must approximately coincide with $\frac{8 T}{\pi}$ at the critical temperature, in contrast to the commonly known $\frac{2 T}{\pi}$ for integer vortices (see Appendix \ref{app:RG}). 
To estimate the critical temperature, we use the Weber-Minnhagen method \cite{WeberMinnhagen}. For each temperature, we fit the data as a function of the system size with the finite-size logarithmic corrections predicted by the second-order renormalization group equations (see Appendix \ref{app:RG}).
We obtain a critical temperature $T_{\text{BKT}}=0.227\pm 0.003 E_J$ for the half-vortex BKT transition, cf. inset of Fig.~\ref{fig:dicehelicity}. Let us note that this result could be slightly modified due to higher-order corrections to the value of helicity-modulus $\rho_s^c$ at the critical temperature.

We provide further evidence for the existence of an unconventional $4e$ low-temperature phase by studying the correlation functions below the critical temperature. The phases of superconductors at low temperatures can indeed be distinguished through the correlation functions of the phase operators $\ee^{in\varphi_{\bf r}}$ \cite{Korshunov1986}, where ${\bf r}$ labels the position of a superconducting island on the lattice and $n=1,2$ indicates whether the operator refers to Cooper pairs or quartets. 
The standard superconducting phase in JJAs is characterized by quasi-long-range order, such that
\begin{equation} \label{C2e}
C_{2e}({\bf r},{\bf r'}) \equiv \left\langle \ee^{i \varphi_{\bf r} }\ee^{-i \varphi_{\bf r'}} \right \rangle \approx \frac{A_{2e}}{\left| {\bf r} - {\bf r'}\right|^\alpha}\,,
\end{equation}
for large distances $\left| {\bf r} - {\bf r'}\right|$, with $\alpha  \approx \sqrt{3}T/(4\pi E_J)$ based on renormalization group approximations valid at low temperature (see Appendix \ref{app:RG}). The previous relation also implies \cite{Kosterlitz1974}:
\begin{equation}\label{C4e}
C_{4e}({\bf r},{\bf r'}) \equiv \left\langle \ee^{i 2\varphi_{\bf r} }\ee^{-i 2\varphi_{\bf r'}} \right \rangle \approx \frac{A_{4e}}{\left| {\bf r} - {\bf r'}\right|^{4\alpha}}\,,
\end{equation}
such that, in the standard superconducting phase, both the phases $\ee^{i\varphi}$ and $\ee^{i2\varphi}$ have correlations that decay asymptotically as a power law, with exponents $\alpha$ and $4 \alpha$ which depend on the system temperature. 
In the previous relations, $A_{2e}$ and $A_{4e}$ are non-universal constants.

The $4e$ phase, instead, is characterized by localized Cooper pairs, corresponding to the fact that the phase $\ee^{i \varphi}$ becomes disordered
\begin{equation}
C_{2e}({\bf{r}},{\bf{r'}}) \approx \tilde{A}\ee^{-\frac{\left| {\bf{r}} - {\bf{r'}}\right|}{\xi_{2e}}}\,,
\end{equation}
where $\xi_{2e}$ is the related correlation length. At $f=1/3$, this exponential decay is caused by the proliferation of zero-energy domain walls as the ones depicted in Fig. \ref{fig:vortex}. Importantly, however, the phase $\ee^{i2\varphi}$ still displays quasi long-range order and fulfills the asymptotic behavior defined by Eq.~\eqref{C4e}. 

Finally, at high temperature, the normal phase displays the exponential decay of all the correlation function of the phases $\ee^{in\varphi}$: no superconducting order parameter can be defined as effect of the deconfinement of fractional vortices.

The contrasting behavior of the correlation functions associated to different harmonics is not limited to the $4e$ phase on the dice lattice, as it emerges, in general, in several physical contexts that can be represented by frustrated XY models. For instance, Lin {\textit{et al.}} showed that the frustrated kagome lattice displays a $6e$-phase characterized by the quasi long-range order of the third harmonics $\ee^{i3\varphi}$ correlations \cite{Song2023,Lin2025} in correspondence with disordered $\ee^{i\varphi}$ and $\ee^{i2\varphi}$ harmonics. The coexistence of correlation functions with exponential and power-law asymptotic decay is indeed one of the signatures of 2D superconducting phases mediated by Cooper multiplets.

To numerically explore the correlations of the superconducting phase at $f=1/3$, we resort to the tensor network representation of the partition function of the XY model \cite{Song2023_general_theory_XY} introduced above. In these calculations, the contraction of the tensor networks is performed approximately using the corner transfer matrix renormalization (CTMRG) algorithm \cite{nishino1996corner_CTMRG1, nishino1997corner_CTMRG2, Corboz2011_CTMRG4.1_unitcell, corboz2014competing_CTMRG5_projectors, Schmoll2021_Heisenberg} with a refinement parameter $\chi_E$ referred to as the environment bond dimension. 
Choosing a finite value for this environment bond dimension induces a maximal correlation length in our numerical calculations,
beyond which correlations will decay exponentially, whatever their true behavior is.

For the sake of simplicity, instead of considering the phase correlation functions defined in Eqs. \eqref{C2e} and \eqref{C4e}, we calculate $\tilde{C}_{2e}(\textbf{r}) = \langle \cos(\varphi_{\textbf{r}} - \varphi_0 )\rangle$ as well as $\tilde{C}_{4e}(\textbf{r}) = \langle \cos(2(\varphi_{\textbf{r}} - \varphi_0 ))\rangle$. 
These are the spin-spin correlation functions of the related XY-model and correspond to the real part of the correlators $C$ discussed above. They are evaluated adopting the Peierls phases $A_{\bf r r'}$ obtained by the fixed gauge choice in Ref. \cite{Rizzi2006}, which realizes a minimal magnetic unit cell.

\begin{figure}[t!]
            \centering
            \includegraphics[width=0.99\linewidth]{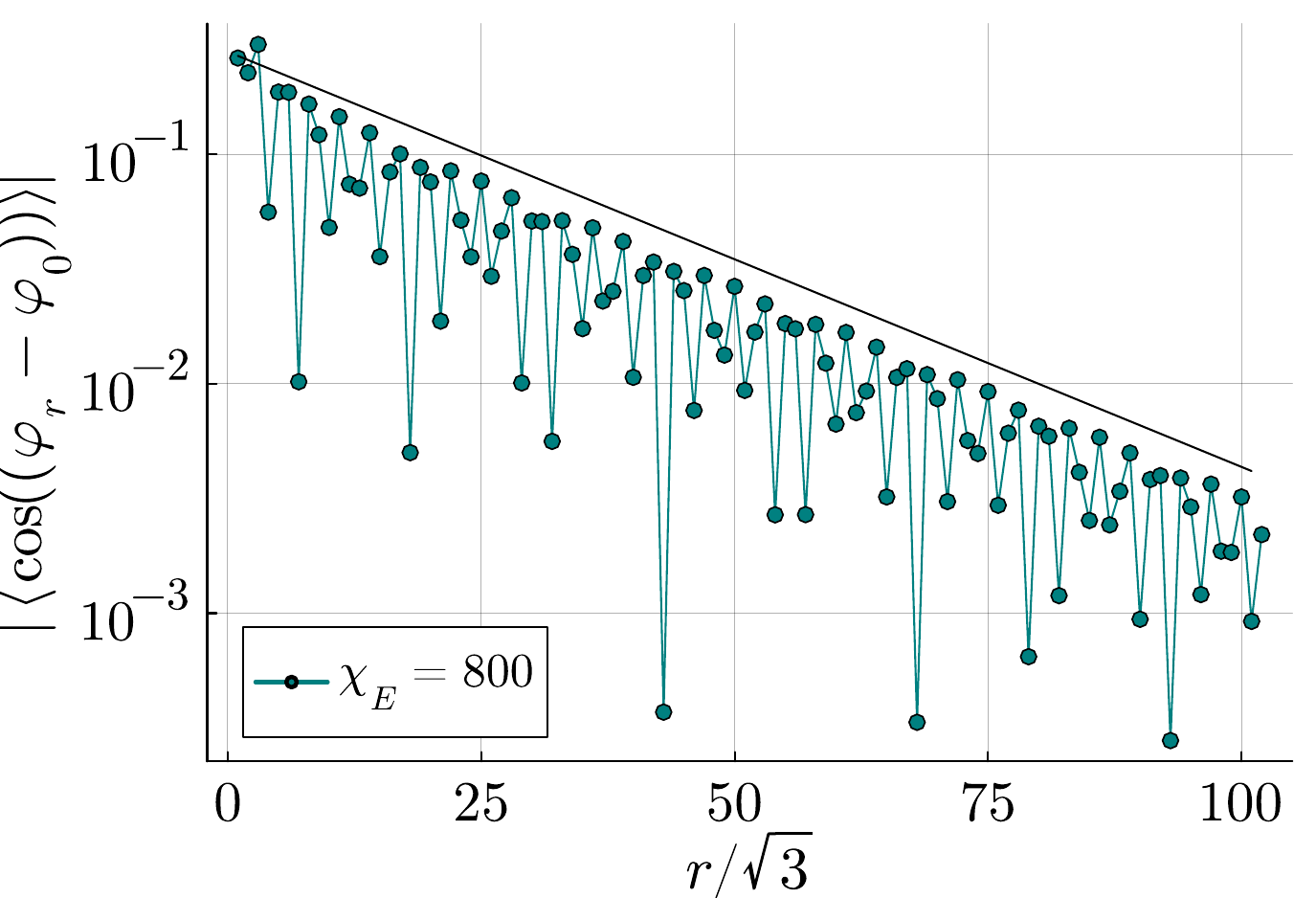}
            \caption{Spin-spin correlation function $\tilde{C}_{2e}(r)$ at $T=0.17 E_J$ as a function of distance along a straight line of hubs of the dice lattice, cf. Fig.~\ref{fig:lattice}. $\tilde{C}_{2e}(r)$ displays a modulated exponential decay.}
            \label{fig:cos}
\end{figure}

Fig.~\ref{fig:cos} shows $\tilde{C}_{2e}(\textbf{r})$ for the temperature $T=0.17 E_J$ below the critical temperature $T_{\rm BKT}$. We observe an exponential decay of $\tilde{C}_{2e}(\textbf{r})$ indicating that no conventional quasi-long-range order is present. 
The oscillating behavior  depends on the gauge choice.

\begin{figure}[t!]
            \centering
            \includegraphics[width=0.99\linewidth]{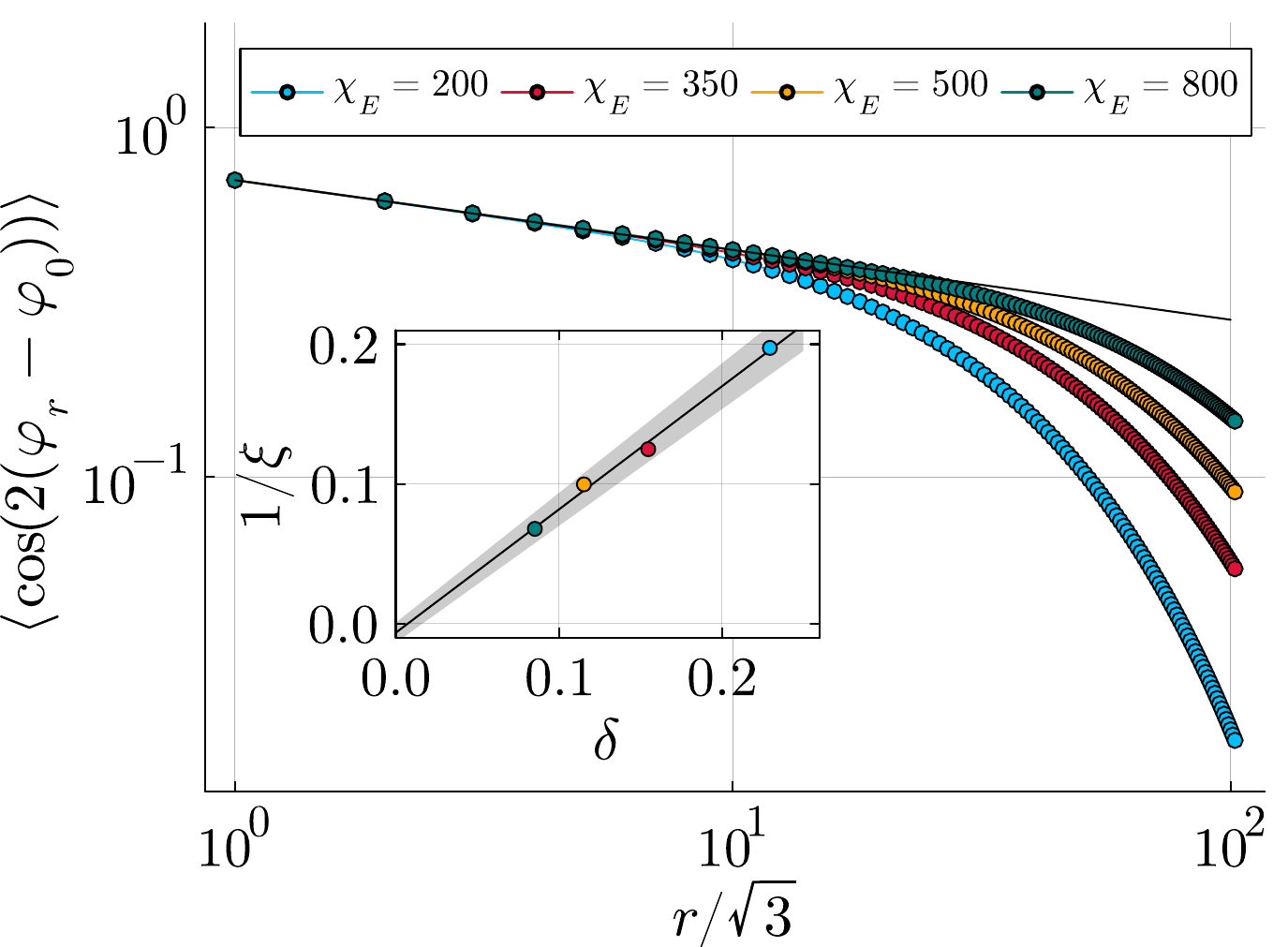}
            \caption{Correlation function $\tilde{C}_{4e}(r)$ at $T=0.17 E_J < T_{\rm BKT}$ as a function of distance along a straight line of hubs of the dice lattice, cf. Fig.~\ref{fig:lattice}. With increasing environment bond dimension $\chi_E$ we find algebraic decay on longer scales before the eventual exponential decay induced by the finite value of $\chi_E$. \textit{Inset:} Extrapolation of the first gap of the transfer matrix (inverse correlation length $1/\xi$) against subleading gap $\delta$ \cite{rams2018precise_scaling_transfermatrix}. The linear fit is consistent with an infinite correlation length.}
            \label{fig:cos2}
\end{figure}

Additionally, we show the correlation function $\tilde{C}_{4e}(\textbf{r})$ at the same temperature in Fig.~\ref{fig:cos2}. In contrast to the case of $\tilde{C}_{2e}(\textbf{r})$, we find all indications of an algebraic decay for this second harmonic correlator. Firstly, we observe that with increasing environment bond dimension $\chi_E$ (the refinement parameter) an algebraic decay persists on increasing distance ranges before, eventually, the exponential decay induced by the $\chi_E$ truncation emerges at large distances as an artifact of our approximation. 
The correlation length $\xi$ is obtained from the (inverse of the) first gap of the spectrum of the transfer matrix of the two-dimensional tensor network. 
In a scenario of algebraically decaying correlations, we expect the spectrum of the transfer matrix to be gapless beyond the first gap as well. We thus expect that a subleading gap $\delta$ will vanish as well in the limit of infinite $\chi_E$.
Extrapolating $\xi$ against $\delta$ for the different values of $\chi_E$ \cite{rams2018precise_scaling_transfermatrix,Vanhecke2019_scaling,Tagliacozzo2008_scaling} (Inset of Fig. \ref{fig:cos2}), we confirm that $\xi$ diverges for $\delta \to 0$. This confirms the algebraic decay of $\tilde{C}_{4e}(\textbf{r})$, as expected in the $4e$ phase below the BKT transition of half-vortices.

This behavior of the correlation functions can be understood from the perspective of the zero-energy domain walls, discussed Sec.~\ref{sec:Ground state properties and half-vortices}. Even at low temperatures, these domain walls proliferate throughout the system. When crossing such domain walls, the superconducting phases on one sublattice shift by $\pi$, cf. Fig.~\ref{fig:vortex}, which destroys the coherence of $\tilde{C}_{2e}(r)$. Contrarily, the correlation function $\tilde{C}_{4e}(r)$ is not impacted by such phase shifts.


\section{The effects of charging energy and quantum fluctuations} \label{sec:charging}

The classical XY model in Eq.~\eqref{eq:2dXYmodel} describes JJAs where the charging energy effects are negligible and all the fluctuations are of thermal nature.
For small superconducting islands and materials with low dielectric constants, however, electrostatic effects become sizeable and, thanks to the tunability of $E_J$, also the ratio of charging and Josephson energies becomes a controllable parameter. 
Therefore, on one side, it is important to assess whether the results discussed so far are stable under the introduction of small charging energies; on the other, it is worth to consider a broader picture and consider which additional phases of matter may emerge in the presence of quantum fluctuations.

In analogous models describing ultracold bosons trapped in optical dice lattices with $f=1/3$, it is known that onsite repulsive interactions lead to the so-called order-by-disorder phenomenon~\cite{Burkov2006}: The introduction of interactions removes the extensive degeneracy of the classical ground state and restores long-range order. 
In particular, Ref. \cite{Burkov2006} discusses the onset of ordered resonating Peierls vortex states which emerge at the third order in perturbation theory over $E_C/E_J$ and favor the honeycomb-like configuration of vortices. Analogous order-by-disorder effects are also observed in the Bose-Hubbard model at $f=1/2$ on the dice lattice \cite{Payrits2014}. 
Additionally, at large $E_C/E_J$, other interesting phases -- compressible but not coherent, like Bose-glasses but without need for disorder~\cite{Rizzi2006} -- could appear, before the system eventually enters the Coulomb-dominated insulating phases.

To estimate the impact of the electrostatic interactions on the $4e$ superfluid phase, we extend the classical two-dimensional XY model in Eq.~\eqref{eq:2dXYmodel} by introducing a local interaction which corresponds to the charging energy of each island, hence obtaining a frustrated quantum phase model \cite{Fazio2001}:
\begin{equation}
H= -E_J \sum_{\left\langle{\bf r},{\bf r'} \right\rangle} \cos \left[\varphi_{\bf r} - \varphi_{\bf r'} -\theta_{\bf r r'} \right] + \frac{E_C}{2} \sum_{\bf r}\left(n_{\bf r} - n_{g,{\bf r}} \right)^2.    \label{eq:full_JJA_Ham}
\end{equation}
Eq. \eqref{eq:full_JJA_Ham} provides a good approximation of the electrostatic interactions for systems with a strong screening, which suppresses inter-island electrostatic interactions, as it is often the case for hybrid systems.  
Here $E_C = 4e^2/C^{\rm self}$, where $C^{\rm self}$ is the island self-capacitance. 

The operator $n_{\bf r}$ acquires integer values and represents the excess number of Cooper pairs on the island $\bf r$. The parameters $n_{g,{\bf r}}$, instead, correspond to the induced charge of each superconducting island and, for the sake of simplicity, we consider the case $n_{g,{\bf r}}=0$. Since we are interested in a scenario with $E_C \ll E_J$, the role of the induced charges and their random fluctuations do not qualitatively affect the physics of the system. The charging energy term induces quantum fluctuations over the classical vortex configurations since phase and number operators, $\varphi_{\bf r}$ and $n_{\bf r}$, are conjugate operators with commutation relation \cite{Fazio2001}
\begin{equation}
    [n_{\bf r}, \varphi_{\bf r'}] = i \delta_{\bf r r'}.
    \label{eq:comm_phase_number}
\end{equation}
The charging energy provides kinetic energy to the vortices and can be interpreted as a sort of disorder operator as it breaks the phase coherence of the islands and reduces the infinite charge fluctuations characterizing the classical states. Therefore these electrostatic effects typically hinder the formation of regular vortex patterns. As we saw in Sec.~\ref{sec:Ground state properties and half-vortices}, however, in the classical $E_C\to 0$ limit of Eq. \eqref{eq:full_JJA_Ham}, the low-energy regime of the dice lattice at frustration $f=1/3$ is characterized by an extensive degeneracy of the ground states associated with different vortex configurations separated by zero-energy domain walls. This extensive degeneracy is not protected by any obvious symmetry, therefore we expect it to be lifted by the quantum fluctuations induced by $E_C$.

We can examine this splitting of the different classical ground states explicitly by considering fluctuations about the corresponding classical configurations. To this end, we separate the phase operators $\varphi_{\bf r} = \varphi^{0}_{\bf r} + \phi_{\bf r}$ into the classical value $\varphi_{\bf r}^0$ and fluctuations $\phi_{\bf r}$. Expanding to second order about the classical ground states, the Hamiltonian in Eq.~\eqref{eq:full_JJA_Ham} can be cast into the quadratic bosonic form of a quantum LC-circuit in momentum space, which can be straightforwardly diagonalized using ladder operators, (see App.~\ref{app:details_Ec} for details).
This allows us to analyse the zero-point energy due to the quantum fluctuations at finite values of $E_C$. This zero-point energy density correction to the classical value of the energy density of a particular vortex configuration is of the form
\begin{equation}
    E_{\text{zp}} = \sqrt{E_CE_J}\Gamma,
\end{equation}
where the coefficient $\Gamma$ involves an integral over all momenta in the Brillouin zone, whose integrand depends on the classical ground state we are expanding about. 

This line of analysis allows us to calculate the zero-point energy correction at small values of $E_C$ for all periodic vortex configurations. 
To get an idea of the different magnitudes of the zero-point energies, we compare the honeycomb vortex pattern, which was already mentioned in Sec .~\ref {sec:Ground state properties and half-vortices} and provides the most stable pattern based for ultracold boson condensates \cite{Burkov2006}, to another canonical vortex configuration, which we refer to as the stripe vortex pattern. The two vortex patterns are illustrated in Fig.~\ref{fig:GS_vortex_configs.} in the Appendix.
We obtain a value of $\Gamma_{\text{h}} = 1.77445/2$ for the honeycomb vortex pattern and a value of $\Gamma_{\text{s}} = 1.77878/2$ for the stripe vortex pattern. 
This suggests that while a finite value of $E_C$ will indeed lift the accidental degeneracy of the classical ground state, for small $E_C/E_J$ the typical difference of the density energy among different vortex patterns can be less than $1\%$ of the plasma frequency $\sqrt{E_J E_C}$ (in the example the difference $\Gamma_{\text{h}} - \Gamma_{\text{s}}$ is in the third significant digit) and cannot be resolved under realistic experimental condition.

Hence, in the limit $T \to 0$ and for a finite $E_C$, we expect the ground state of the model to be characterized by a periodic vortex lattice, corresponding to the phenomenon of order-by-disorder, consistent with Ref.~\cite{Burkov2006}. 

At finite temperature $T$, to minimize the free energy $F = E - TS$, the system will include higher energy states to increase the entropy $S$. 
Given an energy scale $\Delta E_{\textrm{zp}}$ for the splitting of the energy density of the classical ground states at finite $E_C$ and the classical zero-temperature entropy density of the system $S_I$ due to the extensive ground state degeneracy, we can estimate a temperature $T^* = \frac{\Delta E_{\textrm{zp}}}{S_0}$ above which the entropic contribution will dominate the energetic caused by the finite charging energy $E_C$. 

\begin{figure}[t]
    \centering
    \includegraphics[width=0.99\linewidth]{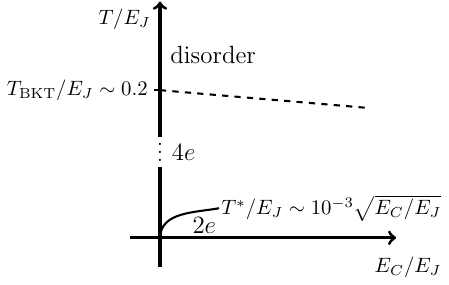}
    \caption{Conjectured phase structure for the dice lattice JJA at $f=1/3$ and small, finite charging energy. The energy splitting due to the quantum fluctuations about the different classical ground states will lift their degeneracy, leading to a vortex lattice phase at $T = 0$. This is associated with conventional 2e quasi-long-range order. At finite temperature, we argue that the the $4e$ phase will reappear at temperatures where the entropic contribution of the many classical ground states overcomes their energy splitting due to the finite value of $E_C$.}
    \label{fig:Phase_diagram}
\end{figure}

Hence, even if at $T=0$ and $E_C>0$ the system lies in a $2e$-ordered phase, we expect that at intermediate temperatures $T^*(E_C) < T < T_{\rm BKT}$ the system will again feature disordered vortex patterns and unconventional quasi-long-range order of the $4e$ kind, as described in Sec. \ref{sec:4ecorr}. 
Taking the assumption that the energy splitting between the honeycomb and stripe vortex pattern described above is a good estimate for the scale $\Delta E_{\textrm{zp}}$ of the energy splitting between the classical ground states, we conjecture the phase diagram structure shown in Fig.~\ref{fig:Phase_diagram} for small values of $E_C$. 

Let us now consider  some typical values of the superconducting array parameters: 
for high transparencies of the junction we set $E_J \sim 0.20\text{meV}$. This value can easily be reduced by depleting the semiconductor through the application of suitable voltage gates. Concerning $E_C$, its value strongly depends on the geometry and chosen materials. Large charging energies have been obtained for small aluminum islands epitaxially grown over two-dimensional InAs heterostructures and may reach typical values of $E_C \sim 0.125\text{meV}$ \cite{Ofarrell2018}. We use this estimate as an indication of the maximum expected charging energy as hybrid JJAs tuned to a high transparency of the Josephson junctions are predicted to present a weaker charge confinement than the Coulomb islands in Ref. \cite{Ofarrell2018}. The resulting temperature $T^*$ separating ordered phase and $4e$ phase is thus $T^*\sim 10 \text{mK}$ and should be compared with the estimated transition temperature $T_{\rm BKT} \sim 0.23 E_J \sim 530 \text{mK}$. 

Besides the introduction of charging energy interactions, the stabilization of an ordered vortex configuration may result also from additional effects. Experiments based on niobium dice networks grown on an insulating substrate observed the onset of the stripe phase for $f=1/3$ \cite{Serret2002}. The main effect that can cause this kind of ordering is given by the mutual inductance of the lattice plaquettes. In the XY model description, this amounts to the introduction of additional classical vortex-vortex interactions beyond nearest neighbor plaquettes which provide an energy cost to the domain walls at $T=0$ \cite{Korshunov2005}. We observe that the plaquette-plaquette inductive coupling roughly scales as $\sim a E_J^2 \mu_0 \Phi_0^{-2} $ and, in JJAs with lattice spacing of order $a \sim 0.5  \text{\textmu m}$, it may display energy scales reaching $1$\textmu eV, thus about $10$mK for large $E_J$. Its effect can therefore be comparable to the energy splitting $\Delta E_{\textrm{zp}}$ such that, at very low temperature $T \lesssim T^*$, induction effects and charging energy may compete to favor different ordered vortex patterns.

\section{Discussion and Outlook}

Our analysis of frustrated JJAs characterized by a dice lattice geometry show that hybrid superconductor-semiconductor architectures provide a controllable and scalable platform for the artificial implementation of a superconducting phase mediated by Cooper quartets. Our results, in particular, suggest that these setups offer a suitable framework for the realization of fractional magnetic vortices, which can be considered the first main ingredient towards the construction of systems with topological order \cite{Ioffe2002,Doucot2003}. A suitable integration of the classical $4e$ superconducting phase with electrostatic interaction effects that provide kinetic energy to the fractional vortices can indeed lead to the formation of topological quantum states. These states would belong to the same universality class of Kitaev's surface code, such that the design of similar JJAs would pave the way for the realization of quantum memories with intrinsic hardware protection.

Based on the frustrated XY model we showed the onset of the 4e superfluid phase for frustration parameter $f=\Phi/\Phi_0=1/3$. In particular, our calculation of the helicity modulus provides a strong indication that the superconductor-insulator phase transition corresponds to the BKT delocalization of half-vortices, and, below the related critical temperature, we observe correlation functions corresponding to a quasi-long-range order for the phase $\ee^{i2\varphi}$ in conjunction with the exponential decay of the $\ee^{i\varphi}$ correlations. This behavior confirms the existence of a $4e$ phase similar to the Lee-Grinstein model, and confirms Korshunov's description of the low temperature behavior based on the proliferation of zero-energy domain walls between ordered vortex patterns \cite{Korshunov2005b}. Additional confirmations of Korshunov's predictions are obtained from the calculation of the extensive entropy of this model at low temperature.


The frustrated XY model that we use in our study corresponds to a description of hybrid JJAs in the regime with strong depletion of the semiconductor rhombic plaquettes \cite{Bondar2025} and the experimental analysis of the JJA resistance above the critical temperature can provide indications about the predicted BKT transition \cite{Newrock2000,Lobb1983,Bottcher2022b}.

Our analysis of the critical current of the model reproduces all the main experimental features at low temperature when accounting for magnetic flux disorder. Furthermore, the harmonics decomposition of the supercurrents obtained by connecting pairs of islands in the bulk via a superconducting loop suggest that Cooper quartets are the main charge carriers for the full range of frustration $1/3 \le f \le 2/3$, such that it is possible that the $4e$ phase extends over a broad range of the magnetic flux per plaquette.

Possible designs for devices able to detect the charge of the current carriers, thus the 4e superfluid phase can be devised based on interferometric setups \cite{Berg2009}, similarly to the recent experiments in kagome superconductor rings \cite{Ge2024}. A rigorous analysis of the critical currents of such devices, however, must account for the random thermally activated phase slips that correspond to the motion of vortices and half-vortices under the insertion of magnetic fluxes within the interferometers. In particular, we expect that the harmonics of the critical current will display both first and second harmonic contributions and their analysis is left for future work.

Additionally, in experimental devices, both electrostatic effects and mutual inductance between the lattice plaquettes may contribute to split the extensive degeneracy of the classical ground states at $f=1/3$. The former cause quantum fluctuations that compete with the extensive degeneracy of the classical limit and can restore the ordering of the vortex patterns based on an order-by-disorder mechanism \cite{Burkov2006,Payrits2014}. 
The latter may favor certain periodic vortex configurations as observed in superconducting network experiments \cite{Serret2002}. Our estimates based on realistic experimental parameters, however, suggest that in both cases these ordering processes will take place at extremely low temperature, below 10mK. Therefore we expect the $4e$ superfluid phase to dominate the physics for a broad range of temperatures, with the possibility to experimentally observe the dynamics of Cooper quartets and half-vortices via transport or scanning SQUID measurements \cite{wang2026}.

\begin{acknowledgments}
We warmly thank L. Banszerus, B. Kalisky, C. Marcus and S. Vaitiek\.enas for sharing their data and for insightful discussions.

This work was partially funded by the Deutsche Forschungsgemeinschaft (DFG, German Research Foundation) via Project-ID 277101999 -- CRC network TRR 183 (``Entangled states of matter'').
E. L. W. thanks the Studienstiftung des deutschen Volkes for support. O.~F.~S. thanks the Niels Bohr Institute Foundation for financial support, and the Condensed Matter Theory group at the Niels Bohr Institute for hospitality. The relaxation simulations were performed on resources provided by Sigma2 - the National Infrastructure for High Performance Computing and Data Storage in Norway, and on the Fox supercomputer at the University of Oslo. The authors gratefully acknowledge the Gauss Centre for Supercomputing e.V. (www.gauss-centre.eu) for funding this project by providing computing time through the John von Neumann Institute for Computing (NIC) on the GCS Supercomputer JUWELS~\cite{JUWELS} (Grant NeTeNeSyQuMa) and the FZ Jülich for computing time on JURECA~\cite{JURECA2021} (institute project PGI-8) at Jülich Supercomputing Centre (JSC).
Part of this work is the content of PhD-thesis of E.L.W. \cite{Thesis_Weerda}.

\end{acknowledgments}


\appendix

\section{Relaxation simulations \label{app:rlxsims}}
\begin{figure}[h!]
  \centering
  \includegraphics[width=0.65\linewidth]{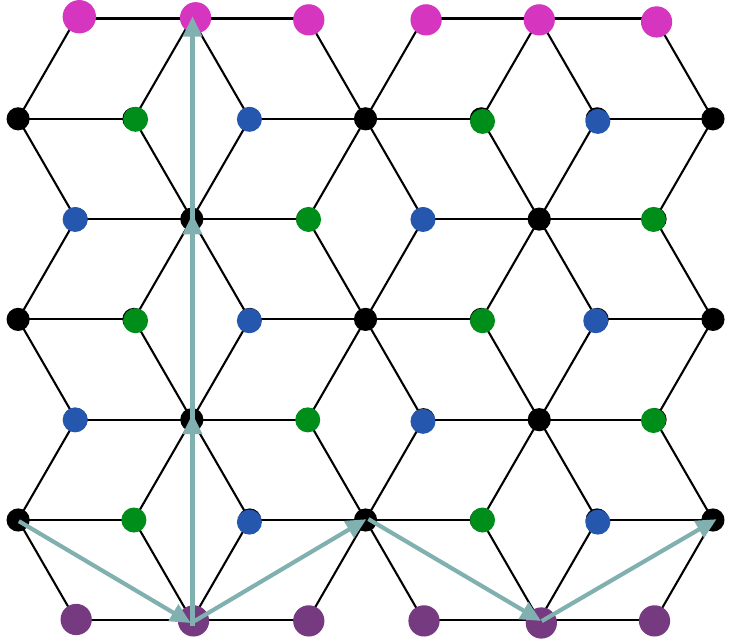}
  \caption{Dice lattice. Superconducting islands (disks) are connected by Josephson junctions (black lines). The islands are divided into three sublattices (black, green, and blue), in addition to source (purple) and drain (pink). The source(drain) islands are always constrained to have identical value of their superconducting phase variable, and are placed on the bottom(top) of the lattice when not otherwise specified. The arrows indicate the extent of the lattice, here $4 \times 3$, in terms of the underlying triangular lattice vectors $\bf e_j$.}
  \label{fig:lattice}
\end{figure}

The relaxation method used here is based on Refs.~\onlinecite{Lobb1983,Straley1988,Halsey1988,Rzchowski1990}.
The phase-biased relaxation simulation starts with the phases in a low temperature state obtained by simulated annealing from a random configuration
using a linear $\beta$-protocol, changing $\beta=0.1$ to $\beta=100$ over $10^6$ Monte Carlo steps (MCS). Each MCS consists of a Metropolis \cite{Metropolis1953} update for every island on each sublattice of the dice lattice, as well as the source and drain islands, performed subsequently, see Fig.~\ref{fig:lattice}.

To ensure a good acceptance ratio, each new phase is proposed to lie in a neighborhood of the old phase \cite{LandauBinder2009,Hinzke1999,Alzate-Cardona2019}.

Once the low temperature state is reached, a further relaxation step (RS) is performed.
In a RS the phases are adjusted so that the local energy of all junctions connected to each island is minimized. 
That is, $\varphi_{\rv}$ on island $\rv$ is selected so that
\begin{equation}
\tan{\varphi_{\rv}} = \f{\sum_{\rvp} \sin{( \varphi_{\rvp} + \theta_{\rv \rvp})}}{\sum_{\rvp} \cos{( \varphi_{\rvp} + \theta_{\rv \rvp})}},
\end{equation}
where the sums are taken over the islands neighboring $\rv$. Besides minimizing energy locally, this also ensures current conservation on island $\rv$. 
This selection is performed sequentially for all islands on a dice sublattice, then one sublattice after the other, and on the source and drain islands, typically hundreds of times until the phases converge.

Then the phase of the drain islands is twisted by an angle $\delta \Phi_{\rm ext}$, which is $2\pi/50$ unless otherwise specified, and a new RS is performed, but this time keeping the phases of the source and drain islands fixed. Once the phases have converged, the supercurrent going into the drain islands is calculated and recorded. This twisting and subsequent RS is repeated to get the current-phase phase relation.
For the global current it is necessary to do many repetitions as many values of $f$ do not show a clear periodic supercurrent-phase relation, thus the (one run) global critical current in Fig.~\ref{Fig:criticalcurrent} is taken as the maximum current within the interval $[0,110 \times 2\pi]$.
For currents between two nearby islands the current-phase relation is regular after an initial transient. Therefore when computing the (local) current harmonics 
\begin{equation}
   I_n = \f{1}{(\Phi_{{\rm ext},2}-\Phi_{{\rm ext},1})} \int_{\Phi_{{\rm ext},1}}^{\Phi_{{\rm ext},2}} \! \! \! \! \! d \Phi \;I_{\rm ext}(\Phi) \ee^{i n \Phi},
\end{equation}
in Figs.~\ref{Fig:Fourier}, \ref{Fig:harmonicshistograms}, \ref{Fig:harmonicshistograms2}, and \ref{Fig:harmonicshistograms3}, we use the interval $[\Phi_{{\rm ext},1},\Phi_{{\rm ext},2}]=[4 \times 2\pi, 7 \times 2\pi]$.  
The whole procedure is repeated, typically 10-100 times, and statistics is recorded. 

\section{Additional data about the two-island tranport at $f=1/3$} \label{app:additionaldata}

In this appendix we present additional data on the joint distribution of the harmonics of the supercurrent flowing between a source and drain island in the bulk of the dice lattice for frustration $f=1/3$.

\begin{figure}[t!]
  \centering
  \includegraphics[width=\linewidth]{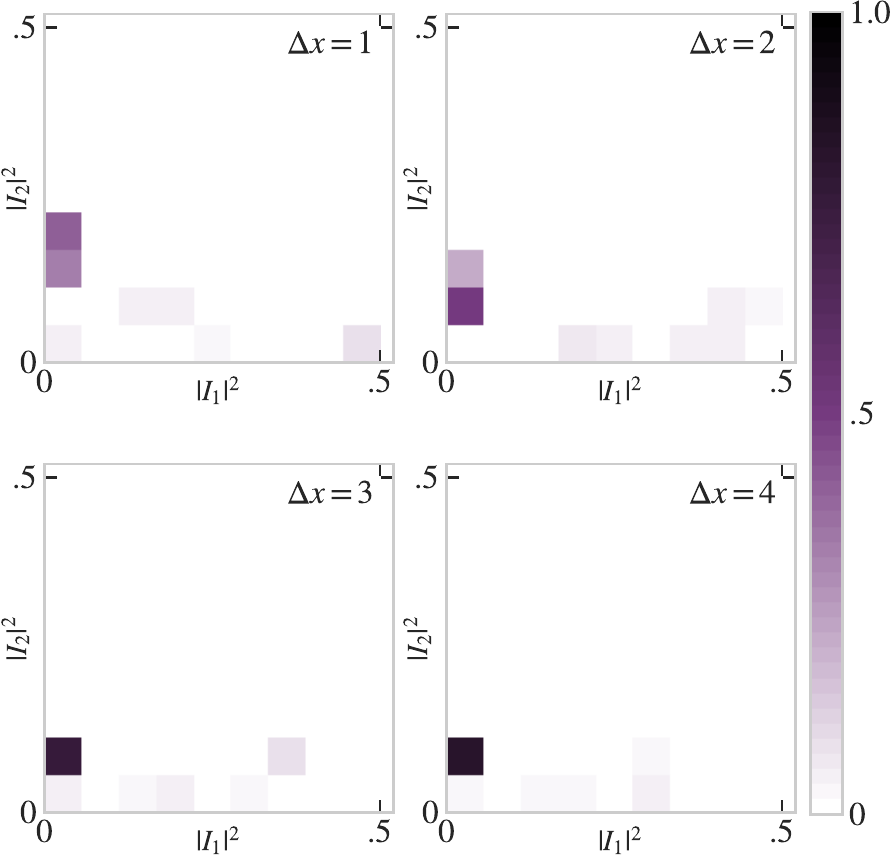}
  \caption{Two-dimensional histograms showing the joint probability of $(|I_1|^2,|I_2|^2)$ at $f=1/3$ for different horizontal distances $\Delta x$ (in units of $3a$) between source and drain located on sixfold coordinated islands near the bulk of a $50 \times 50$ dice lattice. Each histogram is based on 100 independent runs, and the color indicate the fraction of runs within a box.}
  \label{Fig:harmonicshistograms2}
\end{figure}

Fig.~\ref{Fig:harmonicshistograms2} displays the distribution of the $I_1$ Cooper pair and $I_2$ Cooper quartet contribution to the transport between two islands belonging to the sublattice with six-fold coordination. Differently from Fig.~\ref{Fig:harmonicshistograms}, the two islands are displaced along the horizontal direction. Analogously to Fig.~\ref{Fig:harmonicshistograms}, the Cooper quartet contribution ($I_2$) typically dominates over the Cooper pair transport ($I_1$), with a few exceptions likely related to the presence of excitations in the lattice.

\begin{figure}[h!]
  \centering
  \includegraphics[width=\linewidth]{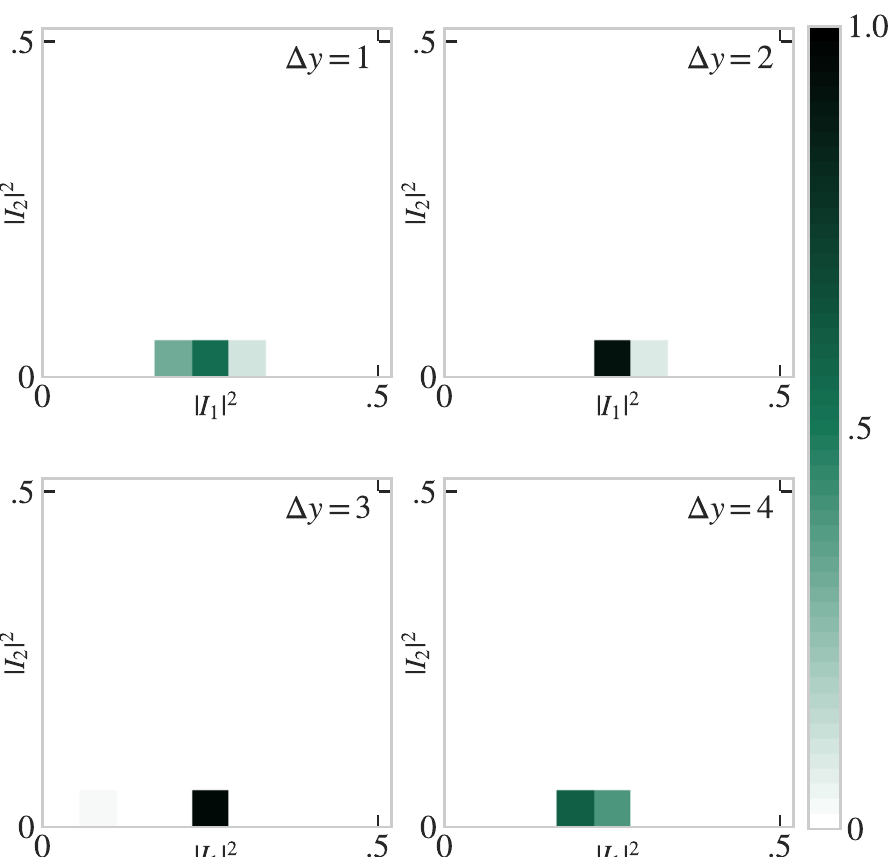}
  \caption{Two-dimensional histograms showing the joint probability of $(|I_1|^2,|I_2|^2)$ at $f=1/3$ for different vertical distances $\Delta y$ (in units of $\sqrt{3}a$) between source and drain located on threefold coordinated islands (green sublattice in Fig.~\ref{fig:lattice}) near the middle of a $50 \times 50$ dice lattice. Each histogram is based on 100 independent runs, and the color indicate the fraction of runs within a box. 
  }
  \label{Fig:harmonicshistograms3}
\end{figure}

The situation is reversed when both source and drain islands are chosen in the three-fold coordination sublattices (see Fig.~\ref{Fig:harmonicshistograms3}): in this case $I_1 \gg I_2$ for all instances.

\section{Renormalization group description of the half-vortex BKT phase transition} \label{app:RG}

In order to describe the low-energy physics of the $4e$ phase, we define an effective action $S=S_0 + S_v$ composed by the slow-oscillating phase contribution $S_0$ and the half-vortex contribution $S_v$.
The effective action $S_0$ can be obtained by considering a suitable gauge-invariant phase difference between the arrays islands and expanding their sinusoidal energy-phase relations. It is given by:
\begin{equation} \label{S0}
S_0 = \frac{K}{2\pi} \int \,d^2 r\, \left(\nabla \varphi \right)^2\,
\end{equation}
where we introduced the effective parameter $K$ whose bare value is $K_0=\frac{2 \pi E_J}{\sqrt{3} T}$, due to the dice geometry. $K$ is related to the helicity modulus of the $XY$ model by $\rho_s = \frac{KT}{\pi} \sim \frac{2E_J}{\sqrt{3}}$, where the last approximation is a rough estimate based on the bare value of $K$.

Based on standard arguments (see, for instance, Refs. \cite{Kosterlitz1974,Newrock2000}), the energy of fractional vortices with vorticity $v_i$ can be expressed, in the continuum limit at low energies, in terms of a Coulomb gas model; the related effective action reads:
\begin{equation}
S_v = -\pi \beta E_J \sum_{i,j} v_i v_j \ln \left( \frac{\left|\bf{r}_i -\bf{r}_j \right|}{a}\right)\,.
\end{equation}
The standard BKT phase transition is retrieved for vorticities $v_i = \pm 1$. In the case of half-vortices, relevant for the $4e$ phase, we consider instead $v_i = \pm v \equiv \pm 1/2$.

By defining $y=\beta E_J$, the related Kosterlitz-Thouless RG equation for half-vortices become:
\begin{equation} \label{RGy}
\frac{dy}{d\ell}=\left(2 - v^2 K \right)y=\left(2 - \frac{K}{4} \right)y\,,
\end{equation}
where $l$ il the RG flow parameter.
The corresponding BKT critical point lies at $K=8$, corresponding to a critical value of the helicity $\rho_s^c = \frac{8T}{\pi}$. The evaluation of the critical temperature obtained by the bare value $K_0$ returns $T_{\rm BKT}\approx 0.45 E_J$, which overestimates the measured critical temperature by about a factor of 2.

This is due to the fact that the renormalization equation \eqref{RGy} is supplemented, at second order, by the renormalization of the $K$ parameter close to criticality:
\begin{equation}
\frac{dK}{dl} = -A y^2\,,
\end{equation}
where $A$ is a non-universal constant.

The corresponding flow towards the critical point at $y=0$ is achieved for $y=\frac{2}{\sqrt{A}}\left(2 - \frac{K}{4} \right)$ and gives:
\begin{equation}
K(l) - 8 = \frac{8}{2l+8 \left({K_0}-8\right)^{-1}}\,.
\end{equation}
For finite-size systems we impose $l=\ln (\tilde{c} L)$, with $\tilde{c}$ a non-universal numerical parameter of order 1; we obtain:
\begin{equation} \label{Weber}
\rho_s(L) - \rho_s^c = \rho_s^c \frac{1}{2\ln L + C}\,,
\end{equation}
with $C \sim 8\left(K_0-8\right)^{-1}$ adopted as a fitting parameter in the Weber-Minnhagen method. This relation provides the finite-size correction estimate adopted to derive the data shown in the inset of Fig. \ref{fig:dicehelicity}.

From the action $S_0$, it is also possible to derive that the critical exponent  in Eq. \eqref{C2e} would correspond to $\alpha=\frac{1}{2K} \approx \frac{\sqrt{3}T}{4\pi E_J}$, such that, in Eq. \eqref{C4e}, one gets the exponent $4\alpha= \frac{2}{K}\approx \frac{\sqrt{3}T}{\pi E_J}$. These estimates are based on the bare parameters and are reliable at low temperature only. When the temperature approaches $T_{\rm BKT}$, the approximation $4\alpha \approx  \frac{\sqrt{3}T}{\pi E_J}$ underestimates the effective decay.

\section{Details on the approximation at finite charging energy
}
\label{app:details_Ec}

By using $\varphi_{\bf r} = \varphi^{0}_{\bf r} + \phi_{\bf r}$ to expand the Hamiltonian in Eq.~\eqref{eq:full_JJA_Ham} to second order about the classical value $\varphi_{\bf r}^0$ in the fluctuations $\phi_{\bf r}$, we obtain
\begin{equation} \label{hamquadr}
    H \approx E_0 + \frac{E_J}{2}\sum_{\langle {\bf i}, {\bf j}\rangle}\cos(\varphi_{ij}^0)(\phi_{\bf j}-\phi_{\bf i})^2 + \frac{E_C}{2}\sum_{\bf i} n_{\bf i}^2,
\end{equation}
with $\varphi_{ij}^0=\varphi^0_{\bf i}-\varphi^0_{\bf j}$ being the classical value of the gauge invariant phase difference on the edge between site $\textbf{i}$ and $\textbf{j}$. Eq. \eqref{hamquadr} can be interpreted as a quantum LC circuit with $C^{\rm self}=4e^2/E_C$ representing the capacitance between each island and the background and $\hbar^2/(4e^2 E_J\cos{\varphi_{ij}^0}) = L_{ij}$ representing the inductance matrix. By Fourier transforming and diagonalizing in the sublattice basis we obtain:
\begin{equation}
    H \approx E_0 - \frac{E_J}{2} \sum_{k,b}\phi_{-k,b}\Gamma(k)_{bb}\phi_{k,b} + \frac{E_C}{2}\sum_{k,b}n_{-k,b}n_{k,b},
\end{equation}
where $b$ is now labeling the bands of the energy dispersion. Reexpressing the Hamiltonian in this form using the operators \begin{align}
\label{eq:x_and_P}
    X_{k,b}^{(1)} = \frac{\phi_{k,b}+ \phi_{-k,b}}{\sqrt{2}},\quad X_{k,b}^{(2)} = \frac{\phi_{k,b}- \phi_{-k,b}}{i\sqrt{2}},\\
    P_{k,b}^{(1)} = \frac{n_{k,b}+ n_{-k,b}}{\sqrt{2}},\quad P_{k,b}^{(2)} = \frac{n_{k,b}- n_{-k,b}}{i\sqrt{2}}
    \label{eq:x_and_P2}
\end{align}
can be used to obtain
\begin{equation} \label{harmo1}
    H \approx E_0 + \sum_{k,b,\sigma}\left[\frac{1}{2m}(P_{k,b}^\sigma)^2 + \frac{1}{2}m\omega_{k,b}^2(X_{k,b}^\sigma)^2\right].
\end{equation}
which has the form of a set of decoupled harmonic oscillators. The index $\sigma \in \{(1),(2)\}$ labels the two different $X$ and $P$ operators introduced in Eq.~\eqref{eq:x_and_P} and \eqref{eq:x_and_P2}. The effective mass and frequency parameters are given by:
\begin{equation}
    m = \frac{2}{E_C},\quad \omega_{k,b}^2 = \frac{-E_JE_C\Gamma(k)_{bb}}{4}.
\end{equation}

Eq. \eqref{harmo1} corresponds to a set of decoupled harmonic oscillators and can be rewritten by introducing suitable ladder operators:
\begin{align}
    a_{k,b,\sigma} = \sqrt{\frac{m\omega_{k,b}}{2}}X_{k,b}^\sigma + i \sqrt{\frac{1}{2 m\omega_{k,b}}}P_{k,b}^\sigma,\\
    a^\dagger_{k,b,\sigma} = \sqrt{\frac{m\omega_{k,b}}{2}}X_{k,b}^\sigma - i \sqrt{\frac{1}{2 m\omega_{k,b}}}P_{k,b}^\sigma.
\end{align}
This allows us to straightforwardly evaluate the zero-point energy of the fluctuations about the ground state $\mathcal{E}_{zp} = \sum_{k,b} \frac{\omega_{k,b}}{2}$.

\begin{figure}[h!]
    \centering
\includegraphics[width=0.49\linewidth]{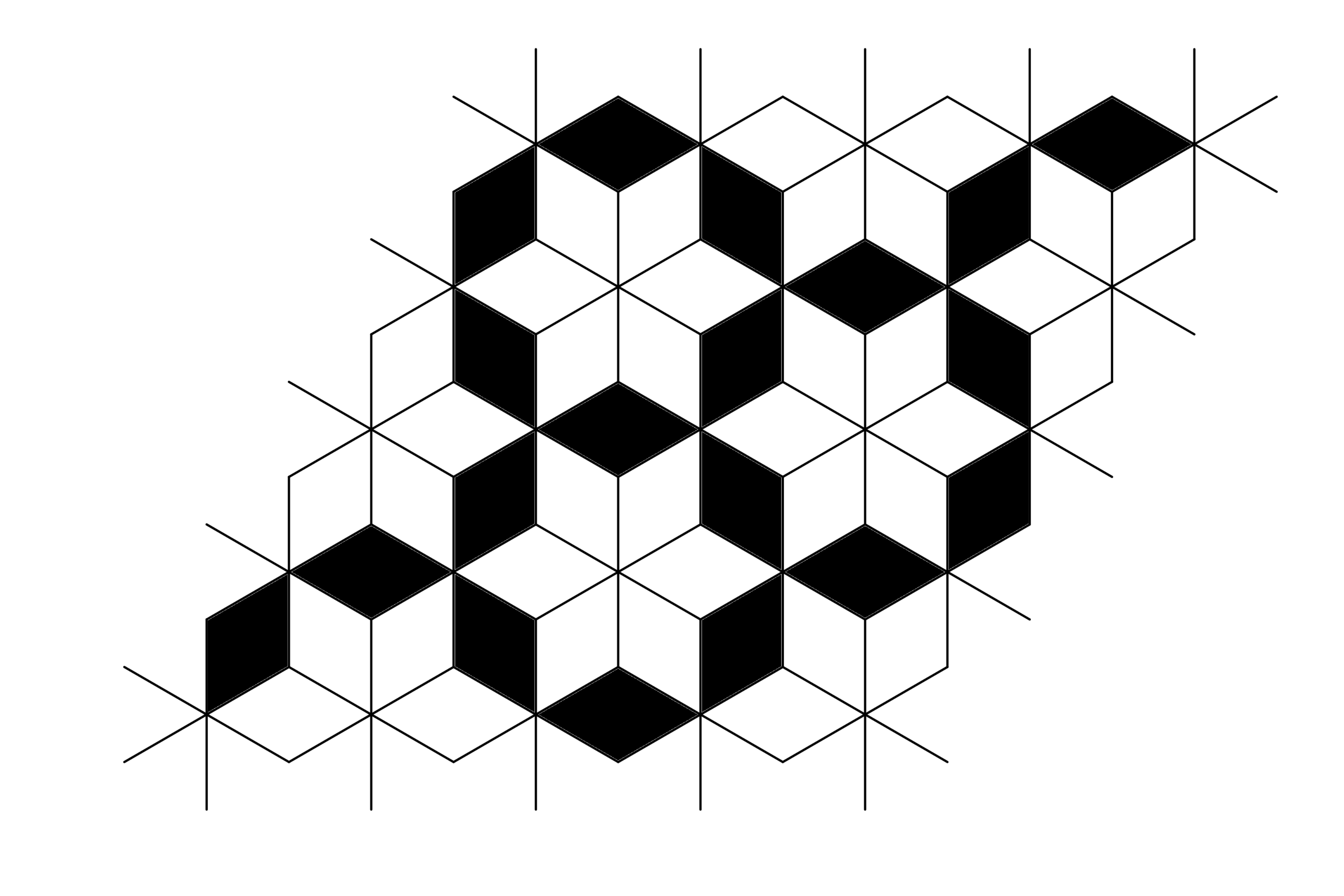}
\includegraphics[width=0.49\linewidth]{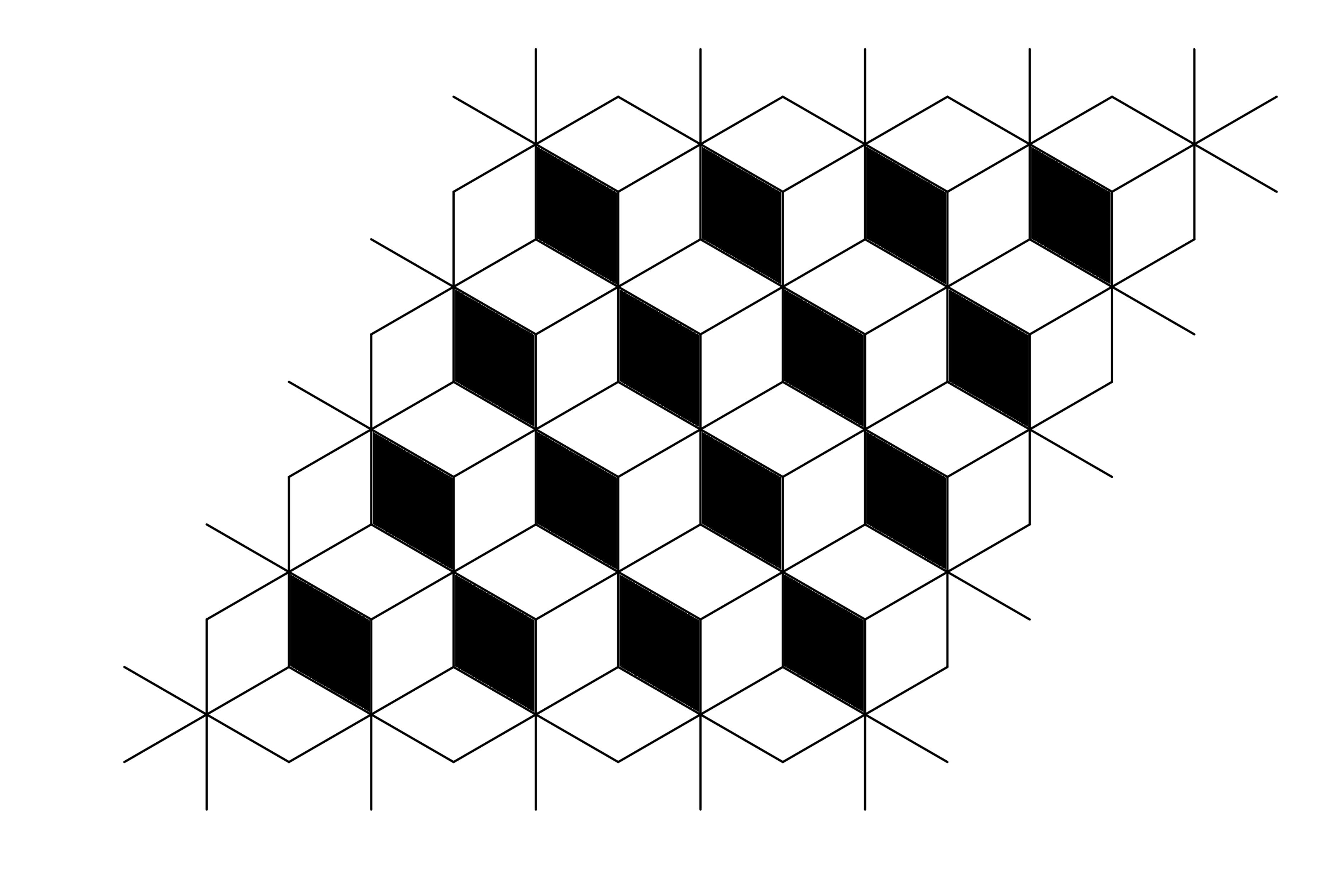}
\caption{Dice lattices with ground state vortex configurations (black rhombi). No two vortices are on adjacent rhombi, and every hexagon of the lattice contains exactly one rhombi. We refer to the configuration on the left as the honeycomb ground state configuration, while we refer to the configuration on the right as the stripe vortex configuration.}
    \label{fig:GS_vortex_configs.}
\end{figure}

Which of the classical ground state configurations identified by Korshunov \cite{Korshunov2005} we start with, will impact the zero-point energy $\mathcal{E}_{zp}$. In the main text, we compared two ground state configurations, which are shown in Fig.~\ref{fig:GS_vortex_configs.}, and we considered the related energy density difference $\Delta E_{zp}$.

\bibliography{biblio.bib}
  
\end{document}